\documentclass{article}
\usepackage{array,latexsym,amsfonts}
\usepackage[dvips]{graphicx,color}
\usepackage{psfrag}
\begin{document}




\title{Self-dual planar hypergraphs and exact bond \\ percolation thresholds} 

\author{John C. Wierman  \\
Johns Hopkins University \\
 and \\
Robert M. Ziff \\
University of Michigan} 



\date{ }

\maketitle

\begin{abstract}
A generalized star-triangle transformation and a concept of
triangle-duality have been introduced recently in the physics
literature to predict exact percolation threshold values of several
lattices.  We investigate the mathematical conditions for the
solution of bond percolation models, and identify an infinite class
of lattice graphs for which exact bond percolation thresholds may be
rigorously determined as the solution of a polynomial equation. This
class is naturally described in terms of hypergraphs, leading to
definitions of planar hypergraphs and self-dual planar hypergraphs.
We show that there exist infinitely many self-dual planar 3-uniform
hypergraphs, and, as a consequence, that there exist infinitely many
real numbers $a \in [0,1]$ for which there are infinitely many
lattices that have bond percolation threshold equal to $a$.
\end{abstract}

\vspace{3mm}

\vspace{3mm}

\section{Introduction}

\subsection{Bond Percolation}

Percolation is an infinite random lattice graph model, which serves
as the simplest example of a process exhibiting a phase transition.
Even so, it provides some extremely challenging problems. Its study
provides intuition for more elaborate statistical mechanics models.
Due to its focus on clustering and connectivity phenomena, it is
applied widely to problems such as magnetism and conductivity in
materials, the spread of epidemics, fluid flow in a random porous
medium, and gelation in polymer systems. Percolation models are
studied extensively in both the mathematical and scientific
literature. See Bollob\'as and Riordan \cite{Bollobas+Riordan},
Grimmett \cite{Grimmett1999}, Hughes \cite{Hughes-volume2}, and
Kesten \cite{Kesten1982} for a comprehensive discussion of
mathematical percolation theory, Stauffer and Aharony
\cite{StaufferAharony1991} for a physical science perspective, and
Sahimi \cite{Sahimi1994} for engineering science applications.

The bond percolation model may be described as follows.  Consider an
infinite connected graph $G$.  Each edge of $G$ is randomly declared
to be open with probability $p$, and otherwise closed, independently
of all other edges, where $0 \le p \le 1$.  The corresponding
parameterized family of product measures on configurations of edges
is denoted by $P_p$.  For each vertex $v \in G$, let $C(v)$ be the
open cluster containing $v$, i.e. the connected component of the
subgraph of open edges in $G$ containing $v$.  Let $|C(v)|$ denote
the number of vertices in $C(v)$.  The percolation threshold of the
bond percolation model on $G$, denoted $p_c(G \mbox{ bond})$, is the
unique real number such that
\begin{equation}
 p>p_c(G \mbox{ bond}) \Longrightarrow P_p (\exists
\hspace{1mm} v \mbox{ such that } |C(v)| = \infty )>0
\end{equation}
and
\begin{equation}
p<p_c(G \mbox{ bond}) \Longrightarrow P_p (\exists \hspace{1mm} v
\mbox{ such that } |C(v)| = \infty )=0 .
\end{equation}

For over fifty years since the origins of percolation theory by
Broadbent and Hammersley \cite{BroadbentHammersley1957}, the
derivation of critical probabilities has been a challenging problem.
Until recently, exact solutions had been proved only for arbitrary
trees \cite{Lyons1990} and a small number of periodic
two-dimensional graphs \cite{Kesten1980, Kesten1982, Wierman1981,
Wierman1984-1525}. These results were obtained using graph duality
and a star-triangle transformation. Scullard
\cite{Scullard-PRE-2006} introduced a generalized star-triangle
transformation which allowed prediction of the exact site
percolation threshold for the so-called ``martini" lattice.  A
triangle-triangle transformation and concept of triangle-duality was
introduced by Ziff \cite{Ziff-PRE-2006} and Chayes and Lei
\cite{Chayes+Lei-JStatPhys-2006}, and further developed by
Ziff and Scullard \cite{Scullard+Ziff-PRE-2006,
Ziff+Scullard-JPhysA-2006}. Triangle-duality allowed
derivation of exact thresholds for an additional collection of
``martini-like" lattices and other lattice graphs.

In this article, we introduce a mathematical framework for unifying
the concepts developed in the previous research. We examine these
new derivations and identify and explain conditions under which the results can be proved rigorously mathematically.  For this purpose,
we describe a class of lattices solvable for the bond
percolation threshold, using the graph-theoretical concept of
hypergraphs, and define planar hypergraphs and a concept of
self-duality for them. We discuss replacing each hyperedge in a
self-dual planar hypergraph by a planar graph called a ``generator"
to obtain a solvable lattice graph.  For the proof that it is
solvable, we construct a dual generator and dual lattice,
and apply the generalized star-triangle transformation to derive the
exact bond percolation threshold.  Certain technical conditions,
such as planarity and periodicity are used to complete a
rigorous mathematical proof of the derivation.  In section 9, we
comment on the possible extension of the method to
site models and nonplanar lattices.

\subsection{The Triangle-Duality Construction}

We first briefly and loosely describe the triangle-duality approach,
in the context of bond percolation, with a slightly different
perspective: We consider constructing a lattice graph rather than
decomposing one. Consider an arrangement of non-overlapping
triangular regions in the plane, with triangles touching only at
their vertices. For convenience, it is sometimes desirable to
represent the triangles as slightly concave, as illustrated in
Figure \ref{self-dual-arrangements}.  Such an arrangement may be
transformed into another arrangement via the ``triangle-triangle
transformation," in which each triangle is replaced by a ``reversed
triangle" as shown in Figure \ref{reversed-triangle}.  If the
resulting (dual) triangular arrangement is equivalent to the
original arrangement, the arrangement is called ``self-dual under
the triangle-triangle transformation" by Ziff and Scullard. If the
triangular arrangement is self-dual, then a lattice may be
constructed by replacing each triangular region by a network of
bonds which has vertices at all three vertices of the triangle. Such
a network will be called the {\it generator} of the lattice. From
such a generator, it may be possible to construct a dual generator,
which creates another lattice when replacing the triangles in the
dual triangular arrangement. By solving an equation derived from the
connection probabilities in the generator and dual generator, a
solution for the percolation threshold is obtained.

One goal of this article is to make
explicit some assumptions which may have been implicit in \cite{Scullard-PRE-2006}, \cite{Scullard+Ziff-PRE-2006}, \cite{Ziff-PRE-2006} and \cite{Ziff+Scullard-JPhysA-2006}. In the remainder of this article, we discuss  conditions which allow a valid exact solution
for the bond percolation threshold in the framework of
3-regular hypergraphs.  Here we only note some remarks and cautions regarding a few issues. (1) Planarity and graph duality play crucial roles in our reasoning, as in all rigorous solutions for bond percolation thresholds of periodic lattices. Our results only directly apply to planar hypergraphs and planar generators.  There is some evidence of wider applicability, which is being investigated. (2) Care must be taken when constructing the
dual hypergraph, with the reversed triangles connected in a precise manner in order to create a proper dual
structure. The reversed triangles need not be the same size or shape as the original triangles, but may need to be distorted instead of merely reversed. (3) In order to prove that the solution is valid, the resulting lattice graph must be periodic. While it is plausible that the results hold for aperiodic models, there is currently no rigorous proof that such solutions are valid.

\subsection{Equality of Percolation Thresholds}

For lattices constructed by this method, the value of the
bond percolation threshold is determined by equations describing the probabilities of connections within the generator.  Therefore, using the same generator in multiple self-dual triangular arrangements produces multiple lattices with equal bond percolation thresholds.  Ziff and
Scullard \cite{Ziff+Scullard-JPhysA-2006}(Figures 1 and 6)
illustrate three different self-dual arrangements. In section
\ref{equality-section}, we show that there are infinitely many
self-dual 3-uniform hypergraphs, so each generator satisfying the appropriate conditions will generate an infinite set of lattices with equal percolation thresholds. Previously, it was only known that there were infinitely many lattices with bond percolation threshold equal to one-half, since Wierman \cite{Wierman-HICStat} provided a construction for infinitely many periodic self-dual lattices.  We also construct a sequence of nested generators which must give unequal percolation thresholds, which implies that there are infinitely
many values $a$ for which there are infinitely many lattices with
bond percolation threshold equal to $a$.  The result also holds for site percolation thresholds, by the bond-to-site transformation.

\section{Background and Definitions}

\subsection{3-Uniform Hypergraphs}

Given a set $V$ of vertices, a hyperedge $H$ is a subset of $V$. A
hyperedge $H$ is said to be incident to each of its vertices.
 A $k$-hyperedge is a hyperedge containing exactly $k$ vertices.  In order
 to neglect the detailed structure of our generators, at times we
 will view a generator as a 3-hyperedge, and will represent
 it in the plane as a shaded triangular region bounded by a slightly concave
 triangular boundary.

 A hypergraph is a vertex set $V$ together with a set of hyperedges
 of vertices in $V$.  A hypergraph containing only $k$-hyperedges is a
 $k$-uniform
 hypergraph. A hypergraph is planar if it
 can be embedded in the plane with each hyperedge represented by a
 bounded region enclosed by a simple closed curve with its vertices
 on the boundary, such that the intersection of two hyperedges is a
 set of vertices in $V$.

In order to construct exactly-solvable lattice graphs for bond
percolation models, we will consider infinite connected planar
periodic 3-uniform hypergraphs. A planar hypergraph $H$ is periodic
if there exists an embedding with a pair of basis vectors {\bf u}
and {\bf v} such that $H$ is invariant under translation by any
integer linear combination of {\bf u} and {\bf v}, and such that
every compact set of the plane is intersected by only finitely many
hyperedges.

If a hypergraph $H$ is planar, we may construct a dual hypergraph
$H^*$ as follows. Place a vertex of $H^*$ in each face of $H$.  For
each hyperedge $e$ of $H$, construct a hyperedge $e^*$ of $H^*$
consisting of the vertices in the faces surrounding $e$. Note that
if the hyperedge is a 3-hyperedge represented by a triangular
region, and each of the boundary vertices is in at least two
hyperedges, then the dual hyperedge is a 3-hyperedge also,
represented by a ``reversed triangle."

Two hypergraphs are isomorphic if there is a one-to-one
correspondence between their vertex sets which preserves all
hyperedges.  A hypergraph is self-dual if it is isomorphic to
its dual.  If the hypergraph is 3-uniform, this corresponds to the
term triangle-dual used by Ziff and Scullard.  To illustrate, in
Figure 1 we provide three examples of infinite connected planar
periodic self-dual 3-uniform hypergraphs mentioned in \cite{Ziff+Scullard-JPhysA-2006}.

\begin{figure}[bhtp]
$$\includegraphics[scale=.40]{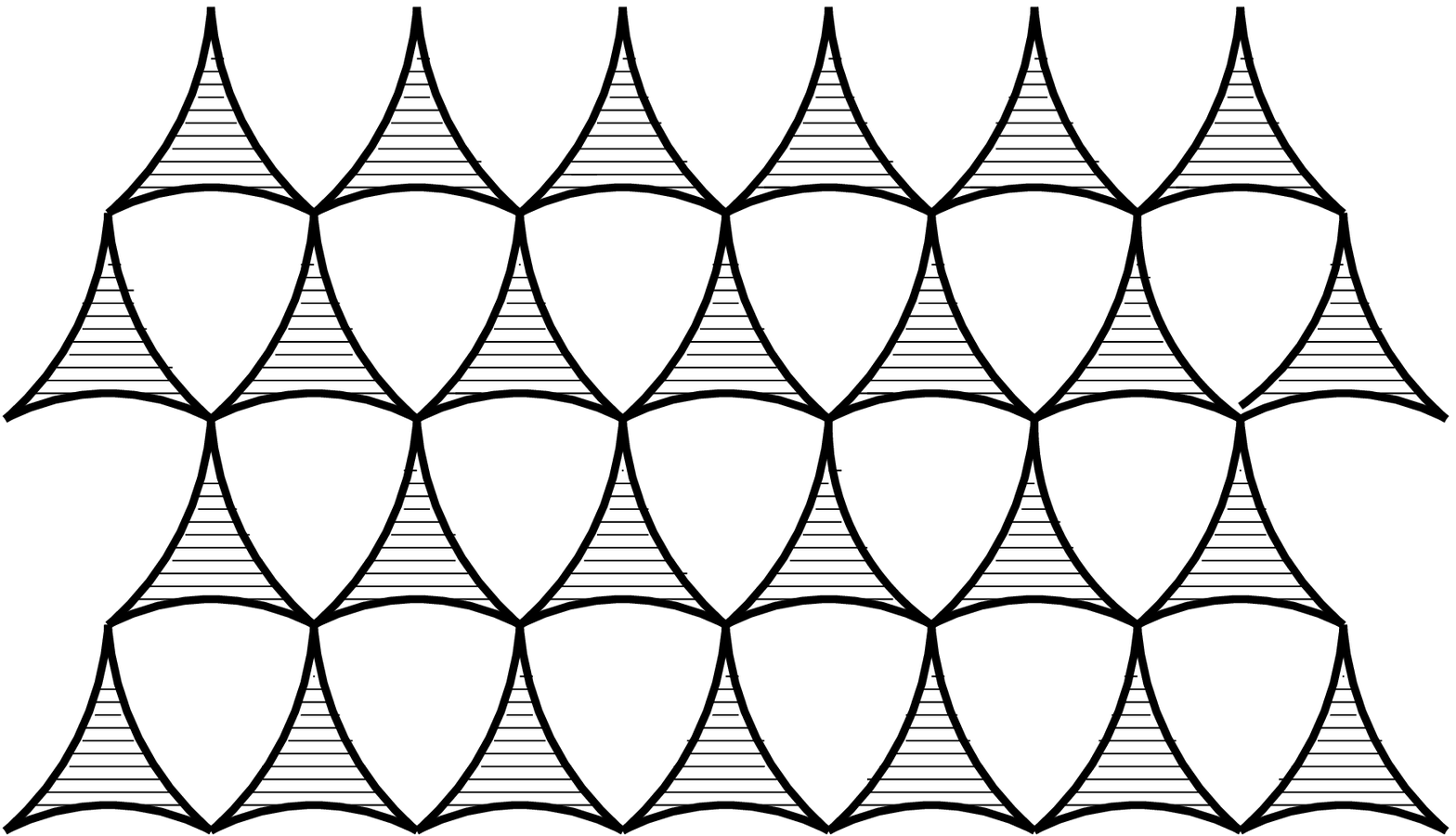}
\hspace{10mm}
\includegraphics[scale=.40]{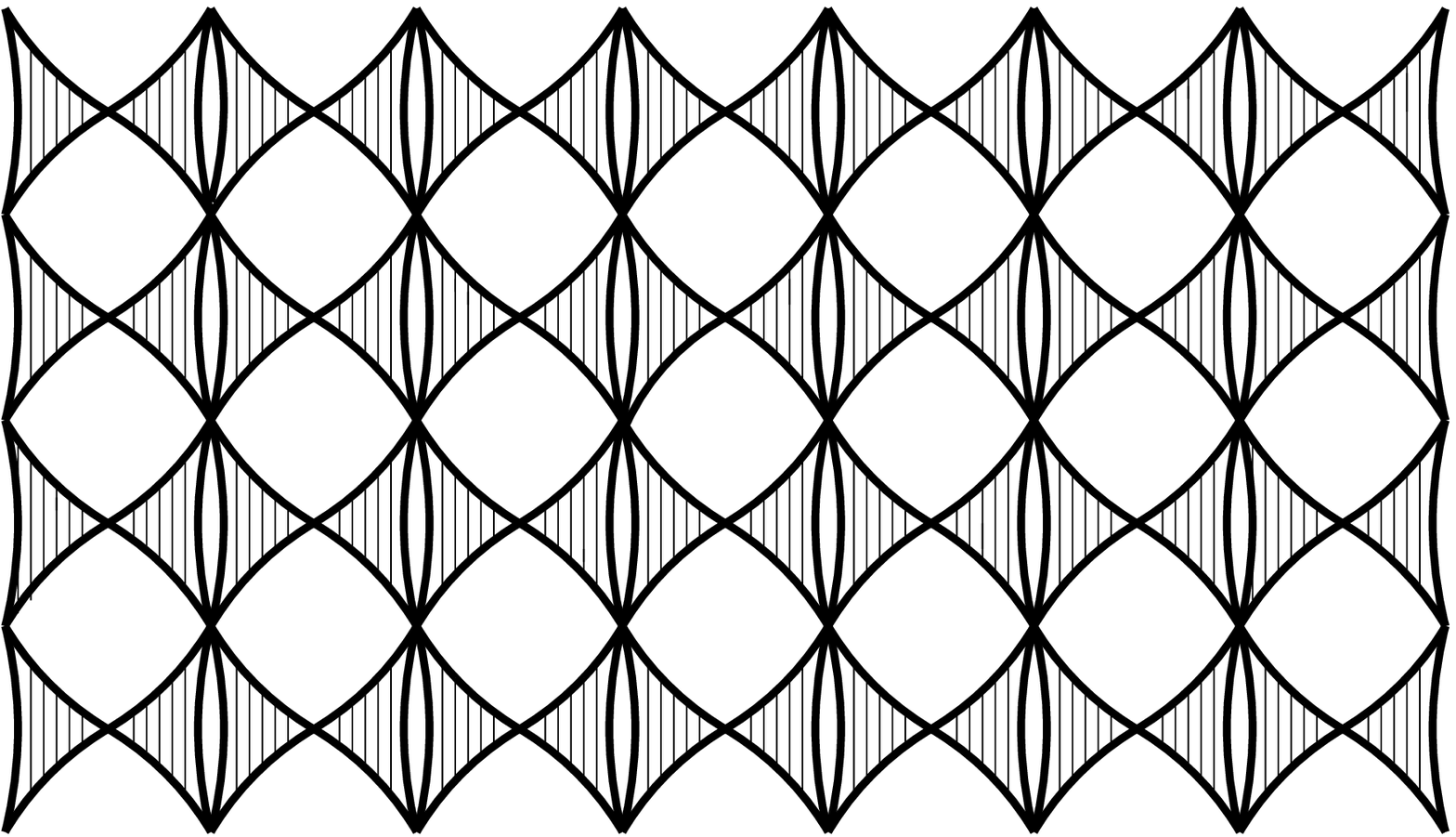}$$
$$\includegraphics[scale=.32]{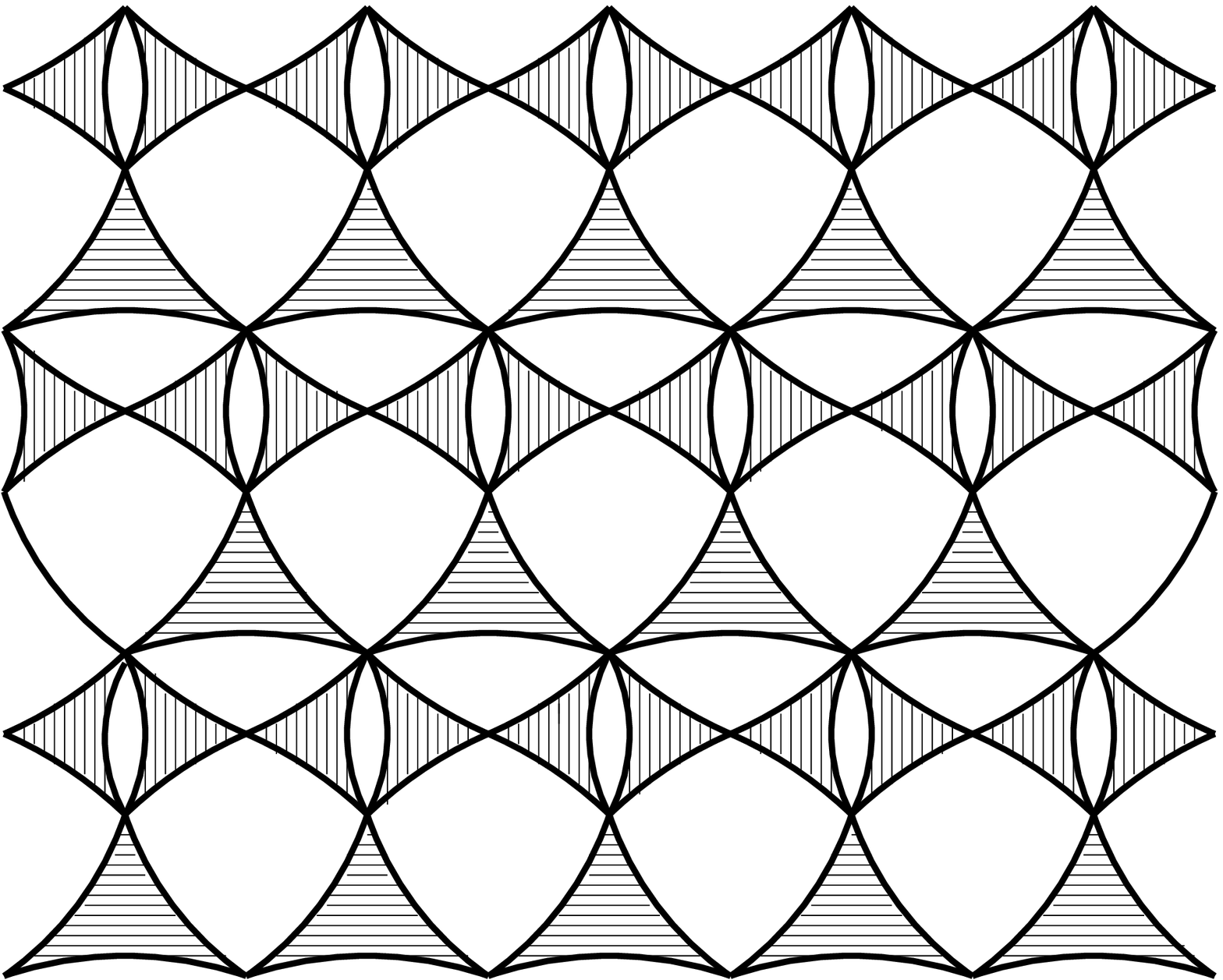}$$
\caption{Self-dual hypergraph arrangements illustrated in \cite{Ziff+Scullard-JPhysA-2006}. In the top row, we refer to the left as the triangular
arrangement, the right as the bow-tie arrangement. The third example
alternates rows of triangles and bow-ties.}
\end{figure} \label{self-dual-arrangements}

As a particular caution, note that the dual hypergraph is not
obtained by simply reversing the triangles in the
original hypergraph. The reversed triangles must be connected in a
specific manner in order to create a proper dual structure.  The way
the reversed triangles are connected in the empty faces of the
original structure is important.  The reversed triangles do not need
to be the same size or shape as the original triangles, but may need
to be distorted instead of simply reversed.  An example of a
hypergraph which appears to be self-dual if one reverses each
triangle, but is actually not self-dual, is given in Figure 2.

\begin{figure}[bhtp]
$$\includegraphics[scale=.30]{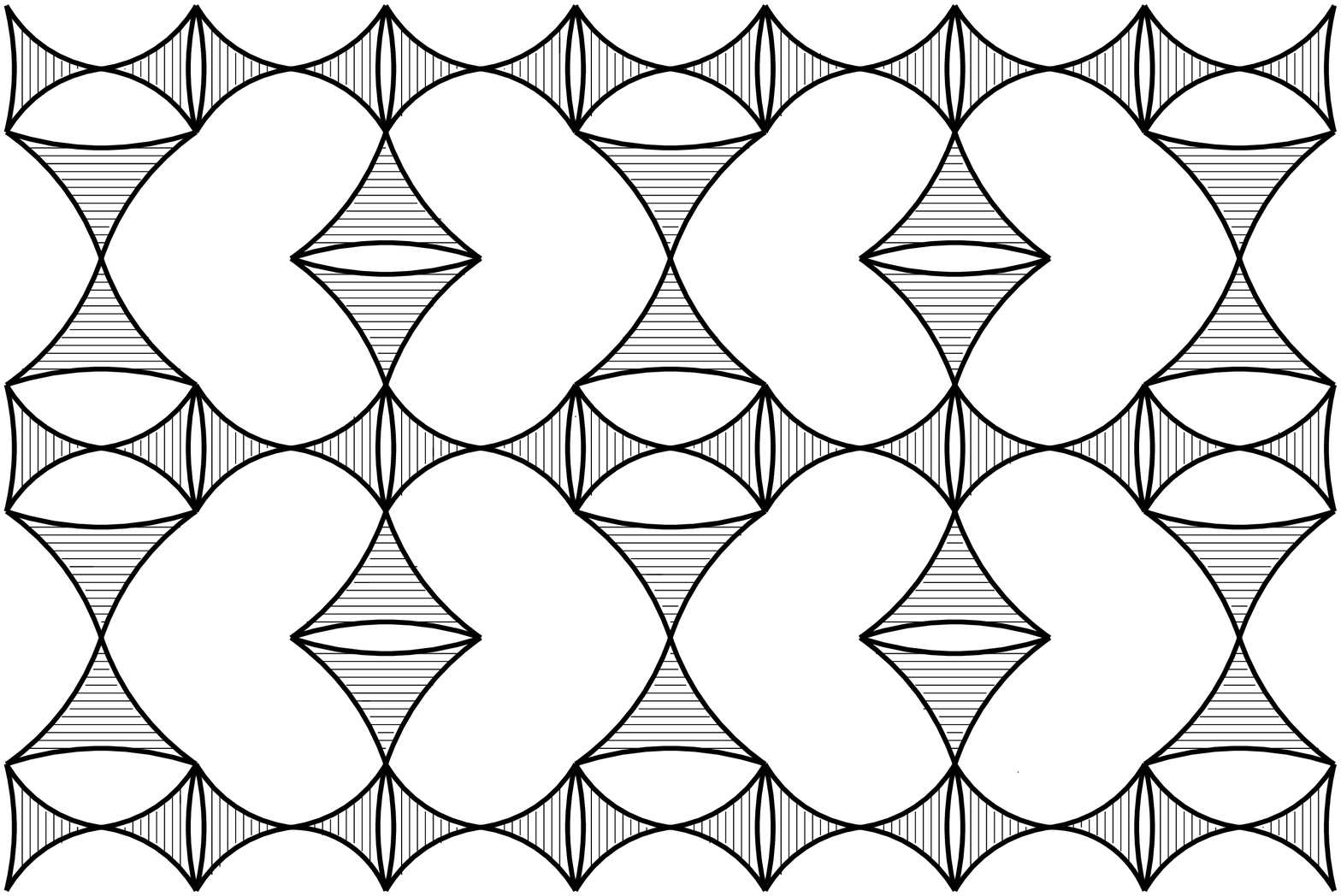}$$
\vspace{5mm}
$$\includegraphics[scale=.30]{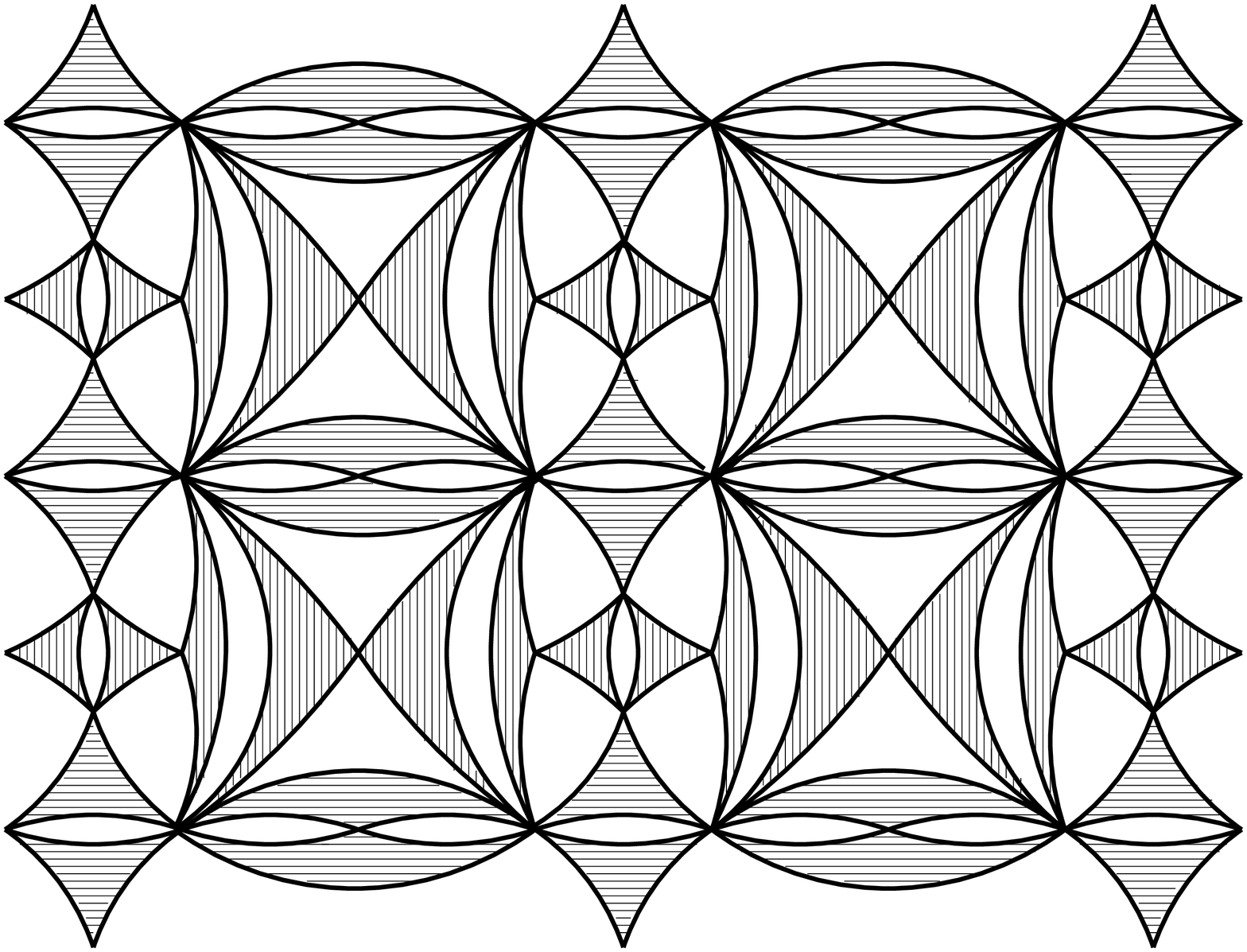}$$  \vspace{5mm}
$$\includegraphics[scale=.30]{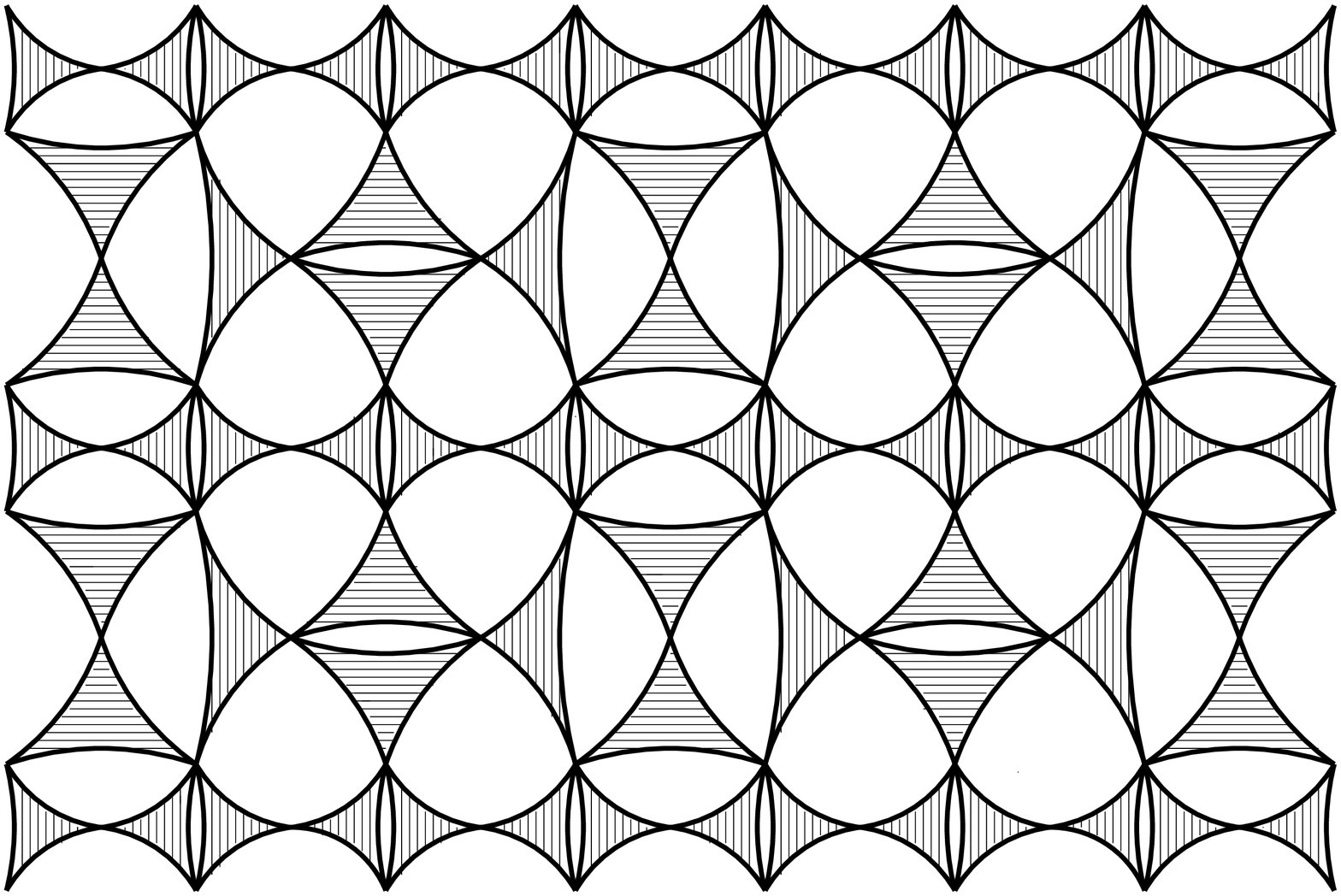}$$

\caption{Top: A example of a hypergraph which is \emph{not} self-dual, but
appears to be if one simply reverses each triangle. Middle: The dual
of the hypergraph above.  Bottom: A self-dual hypergraph constructed
by inserting additional 3-hyperedges in the top arrangement.}
\end{figure}\label{false-example}

\section{Generators and Duality}

A planar graph can be embedded in the plane so that edges meet only
at their endpoints, which divides the plane into regions bounded by
edges, called ``faces."  If the planar graph is finite and
connected, one of these faces is unbounded.  A generator is a finite
connected planar graph embedded in the plane so that three vertices
on the unbounded face are designated as boundary vertices, which we denote as A, B, and
C.

Given a generator G, we construct a dual generator $G^*$ by placing
a vertex in each bounded face of $G$, and three vertices $A^*$,
$B^*$, and $C^*$ of $G^*$ in the unbounded face of $G$, as follows:
The boundary of the unbounded face can be decomposed into three
(possibly intersecting) paths, from $A$ to $B$, $B$ to $C$, and $C$
to $A$. The unbounded face may be partitioned into three unbounded
regions by three non-intersecting polygonal lines starting from $A$,
$B$, and $C$.  Place $A^*$ in the region containing the boundary
path connecting $B$ and $C$, $B^*$ in the region containing the
boundary path connecting $A$ and $C$, and $C^*$ in the region
containing the boundary path connecting $A$ and $B$. $A^*$, $B^*$,
and $C^*$ are the boundary vertices of $G^*$.

For each edge $e$ of $G$, construct an edge $e^*$ of $G^*$ which
crosses $e$ and connects the vertices in the faces on opposite sides
of $e$. If $e$ is on the boundary of the infinite face, connect it
to $A^*$ if $e$ is on the boundary path between $B$ and $C$, to
$B^*$ if $e$ is between $A$ and $C$, and connect it to $C^*$ if $e$
is between $A$ and $B$. (Note that it is possible for $e^*$ to
connect more than one of $A^*$, $B^*$, and $C^*$, for example, if
there is a single edge incident to $A$ in $G$, so its dual edge
connects $B^*$ and $C^*$.)

\begin{figure}[bhtp]
$$\includegraphics[scale=.5]{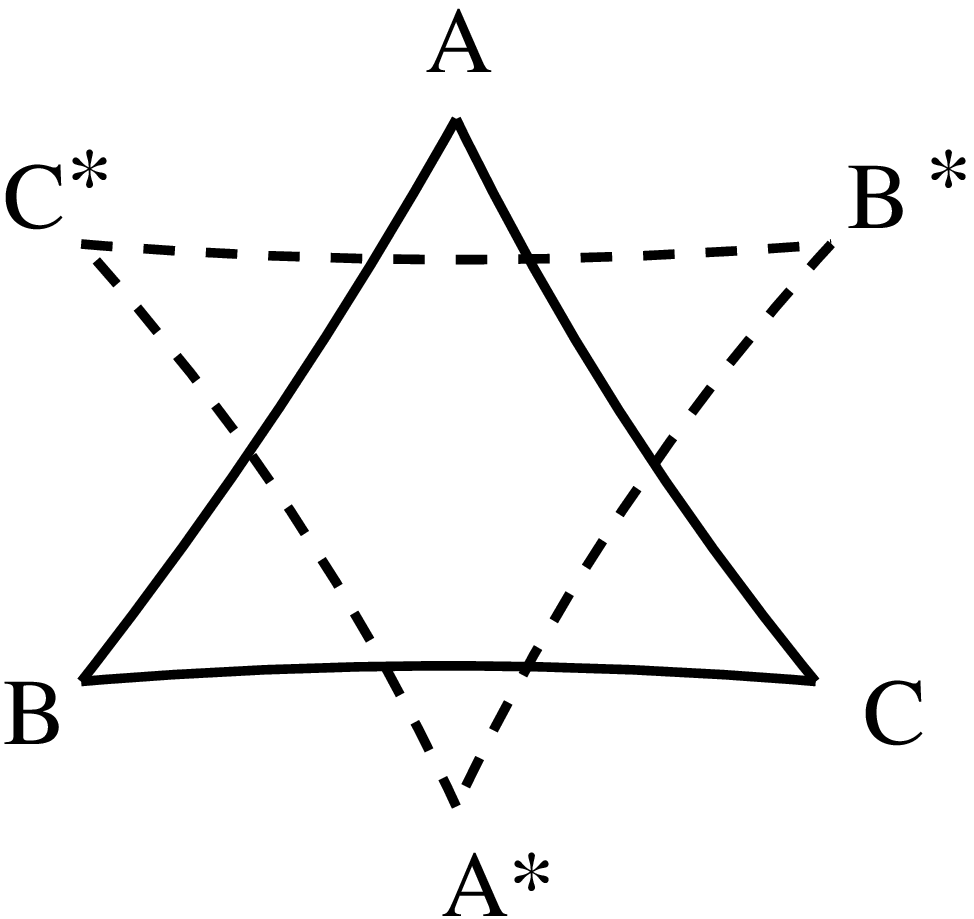}$$
\caption{Solid lines represent a 3-hyperedge with boundary vertices
$A$, $B$, and $C$. Dashed lines represent the ``reversed" or dual
hyperedge, with its boundary vertices $A^*$, $B^*$, and $C^*$
labeled in the proper positions.}
\end{figure} \label{reversed-triangle}

Note that $G^*$ is not the dual graph of $G$, which would have only
one vertex in the unbounded face.  The three vertices $A^*$, $B^*$,
and $C^*$ will correspond to separate faces of the lattice $L_G$
generated from $G$.

Given a planar generator $G$ and a connected periodic self-dual
3-regular hypergraph ${\cal H}$, a dual pair of periodic lattices
may be constructed as follows:  Construct a lattice graph $L_{G,
{\cal H}}$ by replacing each hyperedge of ${\cal H}$ by a copy of the
generator $G$, with the boundary vertices of the generator
corresponding to the vertices of the hyperedge, in such a manner
that the resulting lattice is periodic.  This is always possible, by
choosing the embeddings of the generator in one period of the
hypergraph, and extending the choice periodically.  (However, for a
generator without sufficient symmetry, it may be possible to embed the generator in 
hyperedges in a way that produces a non-periodic lattice, so some
care is needed.)

We now construct a lattice $L_{G^*, {\cal H}^*}$ as follows:
Construct the embedding of the dual hypergraph ${\cal H}^*$ in the
plane, in which every hyperedge of ${\cal H}$ is reversed.  Replace
each hyperedge of ${\cal H}^*$ by a copy of the dual generator
$G^*$, embedded so that it is consistent with the embedding of $G$,
that is, in all hyperedges boundary vertex $A^*$ in $G^*$ is opposite
vertex $A$ in $G$, $B^*$ is opposite $B$, and $C^*$ is opposite $C$,
and each edge of $G^*$ crosses the appropriate edge of $G$. This
results in a simultaneous embedding of $L_{G^*, {\cal H}^*}$ and
$L_{G, {\cal H}}$.  An example of the construction for a particular
generator is illustrated in Figure \ref{construction}.

\begin{figure}[bhtp]
$$\includegraphics[scale=.55]{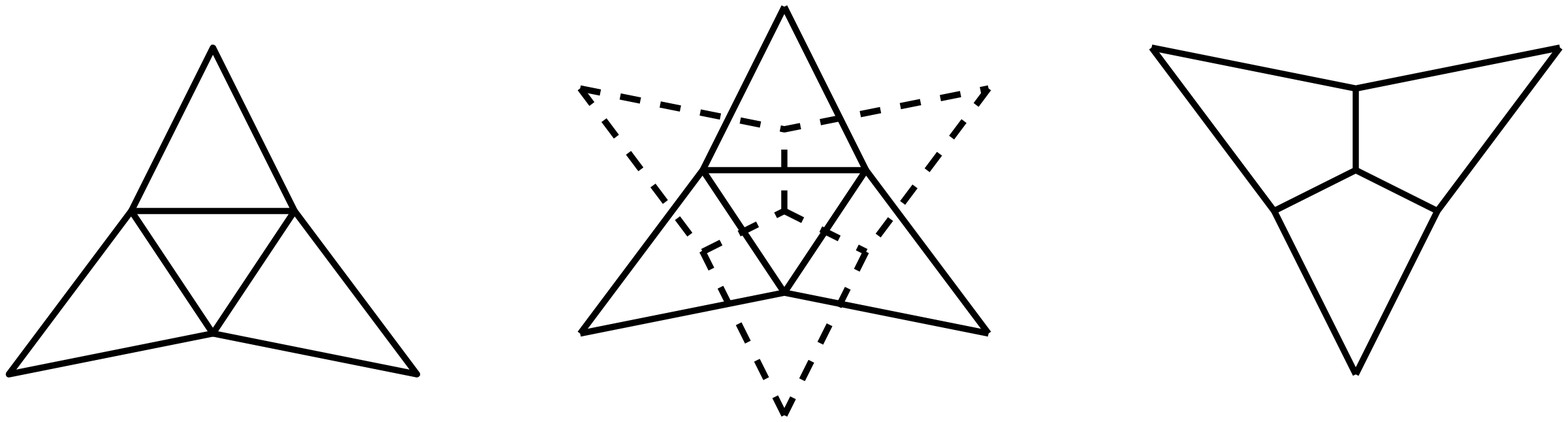}$$
\vspace{5mm}
$$\includegraphics[scale=.25]{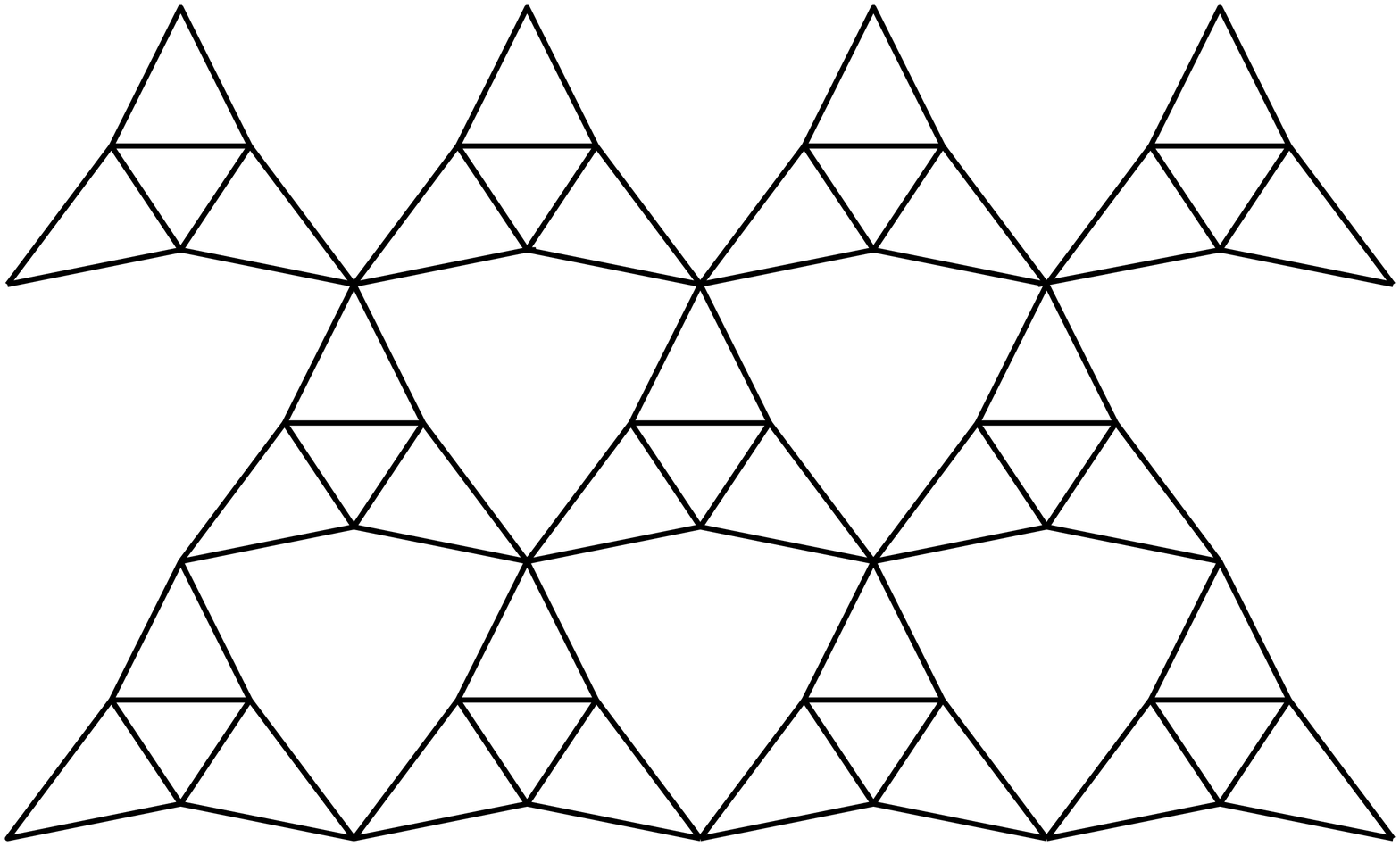}
\hspace{10mm}
\includegraphics[scale=.25]{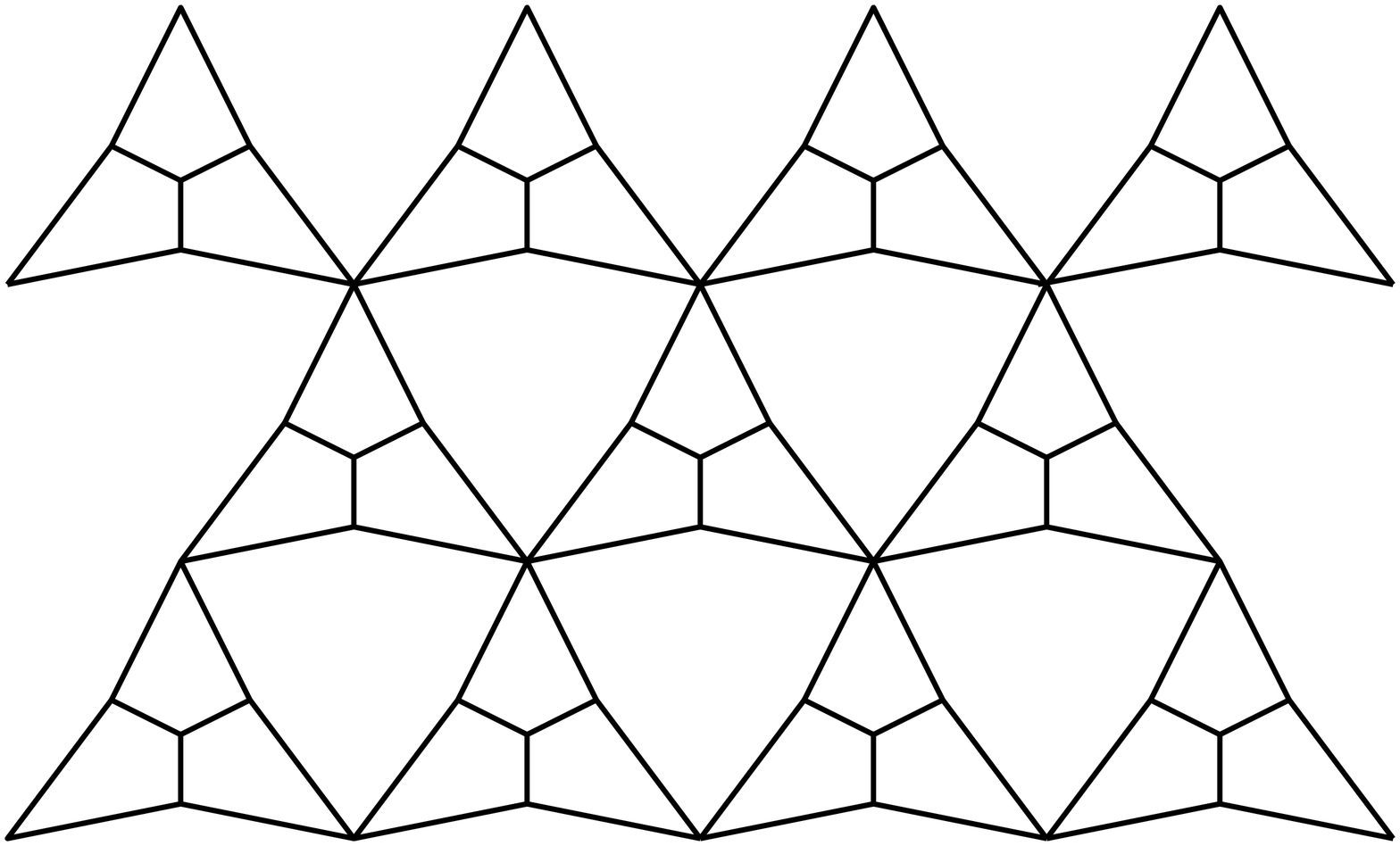}$$
\vspace{5mm}
$$\includegraphics[scale=.25]{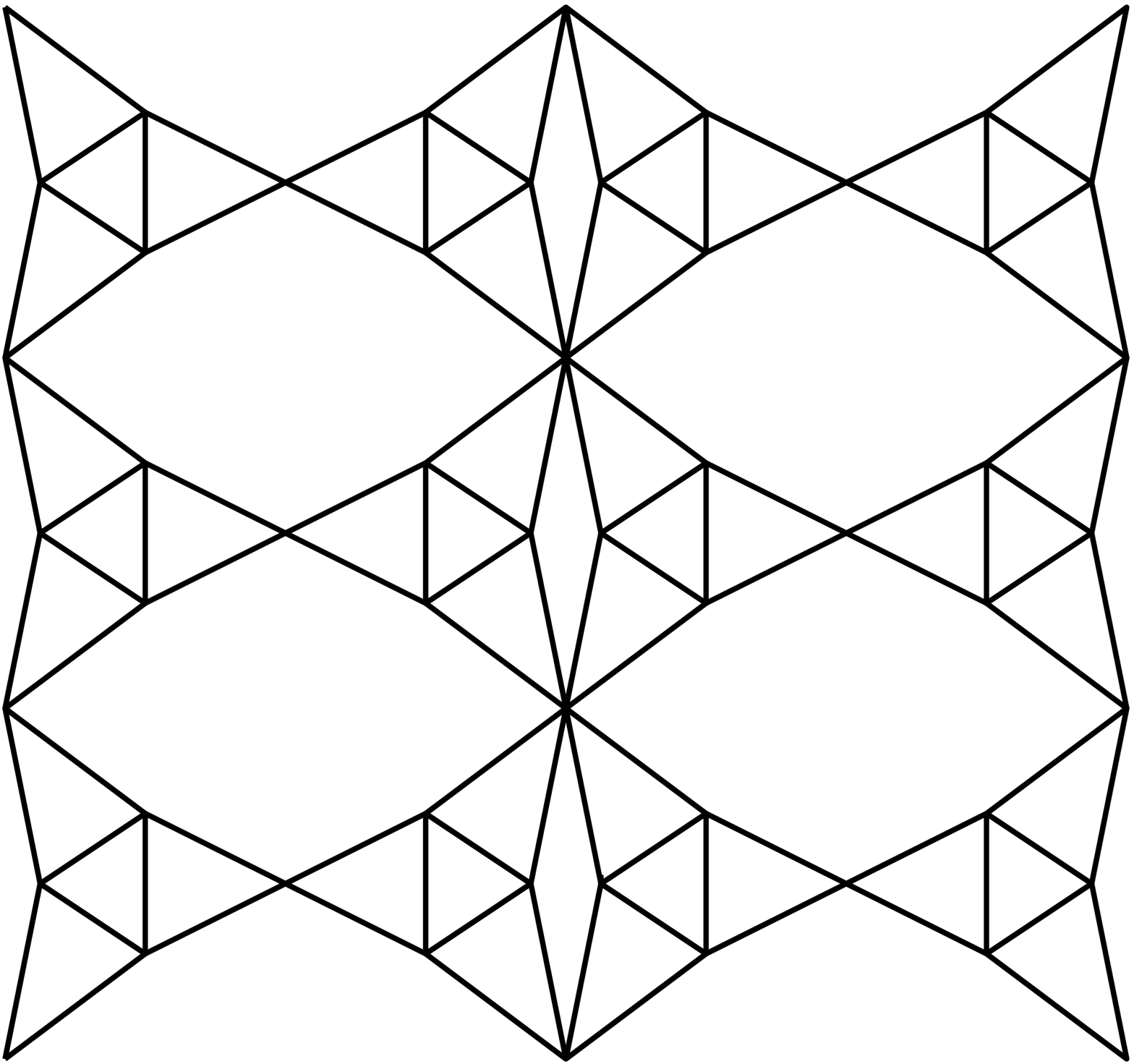}
\hspace{22mm}
\includegraphics[scale=.25]{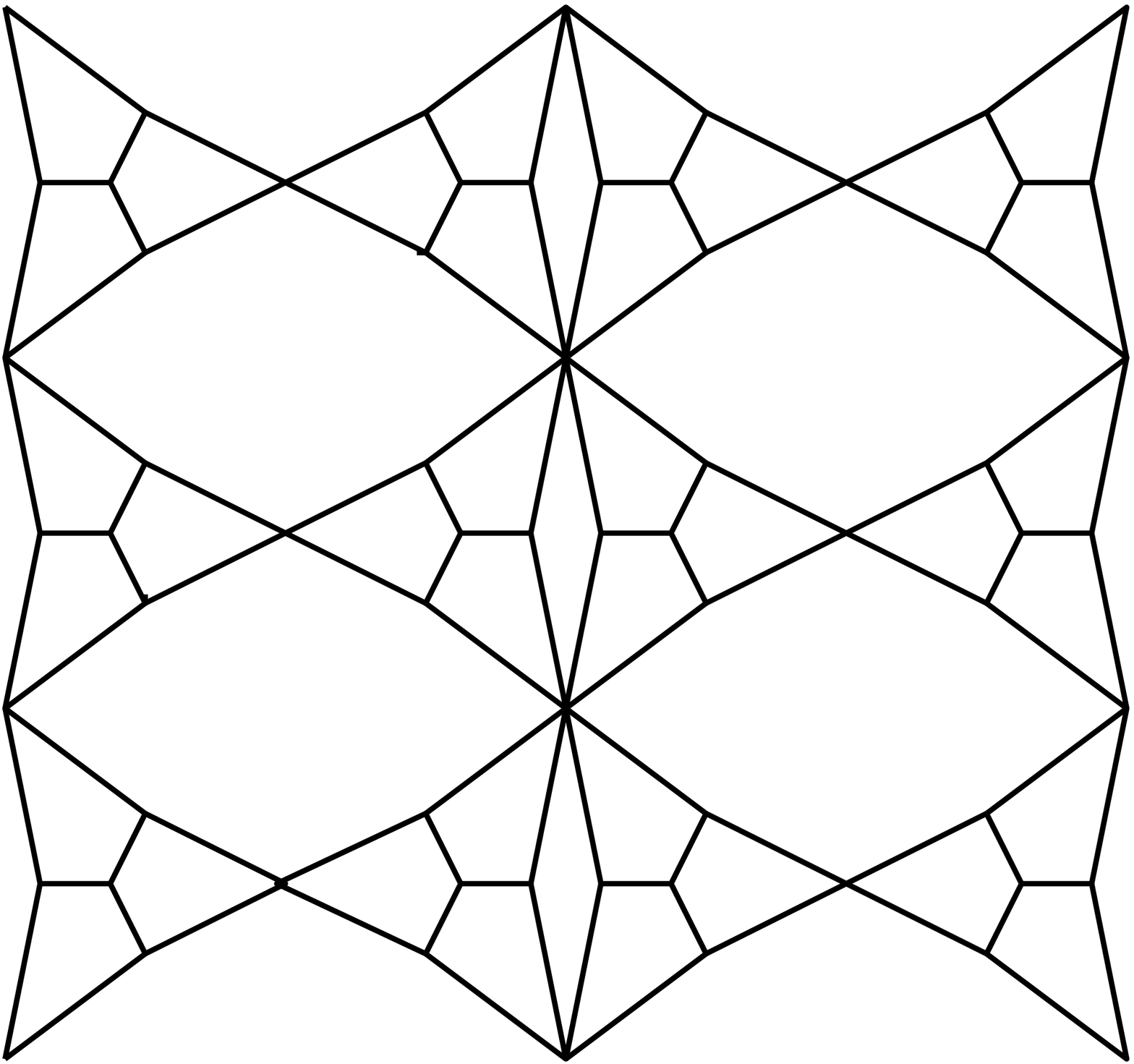}$$
\caption{The construction of lattices based on a specific generator.
Top: The generator, the duality relationship, and the dual
generator. Middle: The lattices based on the generator and the
triangular hypergraph arrangement.  Bottom: The lattices based on
the generator and the bow-tie hypergraph arrangement.}
\end{figure} \label{construction}

The constructions of the two lattices both produce a planar
representation of the resulting lattice. From the simultaneous
embeddings of the two lattices, it is seen that $L_{G^*, {\cal
H}^*}$ is the dual lattice of $L_{G, {\cal H}}$, since there is a
one-to-one correspondence between vertices of one and faces of the
other, and a one-to-one correspondence between edges, which are
paired by crossing. (Note that the position of the boundary vertices
in $L_{G^*, {\cal H}^*}$ is completely determined by the positions
of boundary vertices in $L_{G, {\cal H}}$. Rotations or reflections
of the generator $G^*$ for any hyperedge may not produce a dual pair
of lattices.)

\section{Reduction to a Single Equation}

Consider a generator $G$ and its dual generator $G^*$. In each case,
denote the three boundary vertices by $A$, $B$, and $C$ listed
counterclockwise around the triangle from the initial vertex.  Any
configuration (i.e., designation of edges or vertices as open or
closed) on $G$ determines a partition of the boundary vertices into
clusters of vertices that are connected by open edges. Each such
``boundary partition" may be denoted by a sequence of vertices and
vertical bars, where vertices are in distinct open clusters if and
only if they are separated by a vertical bar.  For example, $AB|C$
indicates that, within $G$, the vertices $A$ and $B$ are in the same
open cluster, but $C$ is in a separate cluster.

Given a planar embedding of the lattice $L_{G, {\cal H}}$ and a
planar embedding of $L_{G^*, {\cal H}^*}$ with each edge crossing
its dual edge, we may define coupled percolation models.  Let each
edge of $L_{G, {\cal H}}$ be open with probability $p$ independently
of all other edges, and define each edge of $L_{G^*, {\cal H}^*}$ to
be open if and only if its dual edge is open.

Suppose we have two bond percolation models on $L_{G, {\cal H}}$ and
$L_{G^*, {\cal H}^*}$, with different edge probability
parameters $p$ and $q$, each assigning probability to configurations
on $G$ and $G^*$, respectively. The probability, denoted $P^G_p (
\pi )$ or $P^{G^*}_q ( \pi )$ respectively, for the partition $\pi$
is determined by summing the probabilities of all configurations
that produce the partition $\pi$ of the boundary vertices.

The set of boundary partitions is a partially ordered set (poset). A
partition $\pi$ is a {\it refinement} of $\sigma$, denoted $\pi \le
\sigma$, if every cluster of $\pi$ is contained entirely in a
cluster of $\sigma$.  The set of boundary partitions ordered by
refinement is a combinatorial lattice, called the partition lattice.

Thus, we have two probability measures, $P^G_p$ and $P^{G^*}_q$ on
the partition lattice, which summarize probabilities of connections between the
boundary vertices without explicitly referring to the detailed
structure of the generator and its dual.  The remarkable fact that
allows exact bond percolation threshold values to be obtained is
that it is possible to choose the parameters $p$ and $q$ so that the
two probability measures are exactly equal.  (Note that in cases
with more boundary vertices, where the probability measures cannot
be made equal, the concept of stochastic ordering of probability
measures may be used to determine mathematically rigorous bounds for
percolation thresholds, using the substitution method
\cite{May+Wierman-2003,May+Wierman-2005,Wierman-1990,Wierman-1995,Wierman-2001,Wierman-2002-RSA,Wierman-2002-CPC}.)

By the duality relationship between $G$ and $G^*$, we have that for
each configuration of open and closed edges, the following five
statements hold:
\begin{enumerate}
\item $A$, $B$, and $C$ are connected by open paths if and only if $A^*$, $B^*$,
and $C^*$ are in separate closed components.
\item $A$ and $B$ are connected by an open path, but $C$ is in a separate
open component if and only if $A^*$ and $B^*$ are connected by a
closed path, but $C^*$ is in a separate closed component.
\item $A$ and $C$ are connected by an open path, but $B$ is in a separate
closed component, if and only if $A^*$ and $C^*$ are connected by a
closed path, but $B^*$ is in a separate closed component.
\item $B$ and $C$ are connected by an open path, but $A$ is in a separate
closed component, if and only if $B^*$ and $C^*$ are connected by a
closed path, but $A^*$ is in a separate closed component.
\item $A$, $B$, and $C$ are in separate open components if and only if
$A^*$, $B^*$, and $C^*$ are connected by closed paths.
\end{enumerate}

While these statements are intuitively clear by drawing diagrams,
the proofs of these statements rely on duality. However, since the
dual generator is not the dual graph of the generator, some
additional vertices and edges must be added to apply graph duality
results. Examples of such reasoning are given in Smythe and Wierman \cite[pp. 8-9]{Smythe+Wierman-1978} and Bollob\'as and Riordan \cite[pp.55-56]{Bollobas+Riordan}.

When considering the random configurations induced by a percolation
model, the five statements become statements of equality of events,
which then have equal probabilities, yielding
\begin{equation}
P^G_p[ABC] = P^{G^*}_{1-p}[A^*|B^*|C^*],
\end{equation}
\begin{equation}
P^G_p[AB|C] = P^{G^*}_{1-p}[A^*B^*|C^*],
\end{equation}
\begin{equation}
P^G_p[AC|B] = P^{G^*}_{1-p}[A^*C^*|B^*],
\end{equation}
\begin{equation}
P^G_p[A|BC] = P^{G^*}_{1-p}[A^*|B^*C^*],
\end{equation}
\begin{equation}
P^G_p[A|B|C] = P^{G^*}_{1-p}[A^*B^*C^*].
\end{equation}

Since $p$ is still a free parameter, we may choose it to satisfy
\begin{equation}
P^G_p[ABC] = P^{G^*}_{1-p}[A^*B^*C^*].
\end{equation}
This equation
always has a solution in [0,1] since the left side is an increasing
polynomial function of $p$ while the right side is decreasing
polynomial, both with values varying between 0 and 1. With this
choice of $p$, the four probabilities in the first and last
equations are equal, and, in fact, the two probability measures are
equal. Thus, it is equivalent to solve the {\it connectivity
equation}
\begin{equation}
P^G_p[ABC] = P^G_p[A|B|C],
\end{equation}
equating the probabilities
of all boundary vertices connected and no boundary vertices
connected in the generator of the lattice. In the following, we
denote the solution to this polynomial equation by $p_0$.

\section{Proof of Exact Solution}

To verify that the solution of equation (9) provides the exact value
of the percolation threshold, we may rely on relatively standard
results in mathematical percolation theory.  Kesten
\cite{Kesten1982} proved that for a dual pair of periodic planar
lattices, $L$ and $L^*$, with at least one axis of symmetry,
\begin{equation}
p_c(L) + p_c(L^*) = 1
\end{equation}
 and that various definitions of
the percolation threshold, denoted by $p_H$, $p_T$, and $p_S$ are
all equal for each lattice. (The result was actually stated for the
more general setting of site models, and applies to bond models by
first applying the bond-to-site transformation to obtain an
equivalent site model.)  To apply Kesten's result to lattices
constructed by the method in this article, we use the $p_S$ concept
of the percolation threshold.

For a periodic graph, the percolation threshold $p_S$ is defined in
terms of the asymptotic behavior of probabilities that open clusters
connect opposite sides of rectangles in sequences of similar
rectangles whose areas are increasing to infinity.  In the
following, we give a brief general description which can be adapted
to each of the lattices in our construction, but avoiding some
technicalities that are specific to each lattice.

Now let $L$ be a lattice constructed in this article, and let $L^*$
denote its dual lattice. $L$ may be embedded periodically in the
plane such that the unit vectors in the $x$- and $y$-axis directions
are the basis vectors and the $x$-axis is an axis of symmetry. For
each pair of positive integers, $m$ and $n$, let $L_{m \times n}$
denote the graph containing all edges of hyperedges of $L$ that
intersect the region $[0,m] \times [0,n]$. We can specify sets of
vertices of $L_{m \times n}$ which we call its top, bottom, left,
and right sides.  We are interested in the event that there is an
open path which crosses $L_{m \times n}$ from left to right, or top
to bottom, which we denote by $\{ open \leftrightarrow L_{m \times
n} \}$, or $\{ open \updownarrow L_{m \times n} \}$, respectively.
Similarly, we consider also closed crossings. The percolation
threshold $p_S$ is defined as
\begin{equation}
\hspace{-16mm} p_S = \sup \{ p: \lim_{n \rightarrow \infty} P_p
[open \updownarrow L_{3n \times n}]=0 \mbox{ and } \lim_{n
\rightarrow \infty} P_p [open \leftrightarrow L_{n \times 3n}]=0 \}.
\end{equation}

In the following, we will use square regions, noting that if
$\limsup_{n \rightarrow \infty} P_p[ open \leftrightarrow L_{n
\times n}] > 0$, then $\limsup_{n \rightarrow \infty} P_p[ open
\leftrightarrow L_{n \times 3n}] > 0$, and similarly for vertical
crossings. In either case, $p > p_S.$

Now let $L_{n \times m}^*$ denote the graph corresponding to the
dual hyperedges of the hyperedges in $L_{n \times m}$.  For $L_{n
\times m}^*$, we can also specify sets of vertices of the lattices
which we can call the top, bottom, left, and right sides of these
regions, and define similar crossing events, in such a manner that
duality yields
\begin{equation}
P_{p_0} [ open \leftrightarrow  \mbox{ in } L_{n \times n}] +
P_{p_0} [ closed \updownarrow  \mbox{ in } L^*_{n \times n} ] = 1,
\end{equation}
which implies that
\begin{equation}
P_{p_0} [ open \leftrightarrow  \mbox{ in } L_{n \times n}] +
P_{1-p_0} [ open \updownarrow  \mbox{ in } L^*_{n \times n} ] = 1.
\end{equation}

We now replace generators of $L^*$ by generators of $L$. In the
duality relationship, each hyperedge in $L^*$ is obtained by
rotating the corresponding hyperedge of $L$ by $180^o$ degrees, so
this replacement results in a copy of $L$ reflected through some
horizontal line.  However, the subgraph of $L$ corresponding to the
replacement of hyperedges in $L^*_{n \times n}$ may not be
isomorphic to $L_{n \times n}$.  For convenience, the following
argument is written as if it is isomorphic.  (If not, by
monotonicity, we may consider a rectangle which is slightly smaller
vertically and slightly larger horizontally, which is sufficient to
obtain the conclusion.)

While replacing the hyperedges, by using the equality of
connectivities at parameters $p_0$ in $L$ and $1-p_0$ in $L^*$, we
have
\begin{equation}
 P_{p_0} [ open \leftrightarrow  \mbox{ in } L_{n \times n}] +
P_{p_0} [ open \updownarrow \mbox{ in } L_{n \times n} ] = 1.
\end{equation}

As a consequence, either
\begin{equation}
\limsup_{n \rightarrow \infty}  P_{p_0} [ open \leftrightarrow
\mbox{ in } L_{n \times n}] \geq \frac{1}{2}
\end{equation}
or
\begin{equation}
 \limsup_{n \rightarrow \infty} P_{p_0} [ open
\updownarrow \mbox{ in } L_{n \times n} ] \geq \frac{1}{2},
\end{equation}
 (or both) so
\begin{equation}
p_S (L) \leq p_0.
\end{equation}

Reversing the roles of $L$ and $L^*$, we also obtain
\begin{equation}
p_S (L^*) \leq 1 - p_0.
\end{equation}

Therefore, \begin{equation}
 1 \leq p_c(L)+p_c(L^*) = p_S(L)+p_S(L^*)
\leq p_0 + (1-p_0) = 1,
\end{equation}
 so the percolation thresholds must equal
their upper bounds, yielding
\begin{equation}
p_c(L) = p_0
\end{equation}
 and
\begin{equation}
p_c(L^*)= 1 - p_0.
\end{equation}

Note that, for periodic lattices that do not have an axis of reflection symmetry, there is no
general proof that $p_c(L) + p_c(L^*) = 1,$ in which case the
reasoning above is not valid.

\section{Subdivided and Split Edges}

There are two situations which can produce lattices which do not occur in natural physical models.  If the generator has a
boundary vertex with only one incident edge, it is possible for the
resulting lattice to have two edges in series. If the generator has
an edge which is incident to two boundary vertices, it is possible
for the resulting lattice to have double edges, i.e., two edges
between the same pair of vertices.

These problems cannot occur for some self-dual hypergraphs:  In any
hypergraph for which each boundary vertex is shared by three
or more hyperedges, such as the triangular lattice arrangement, each
boundary vertex has at least three incident edges, and thus there are
no edges in series. In any hypergraph in which any two hyperedges
intersect in at most one boundary vertex, such as the triangular
arrangement, there cannot be parallel edges.

The generators illustrated in Figures \ref{construction} and
\ref{nested-generators} do not produce subdivided or split edges in
any self-dual hypergraph arrangement. However, for other hypergraph arrangements, such as the bow-tie
hypergraph, there can be subdivided edges and split edges.  If all
hyperedges have the same number of pendant boundary vertices
involved in series edges and the same number of adjacent boundary
vertices involved in parallel edges, in a periodic pattern, then the
lattice can be transformed into a lattice without series or parallel
edges for which the bond percolation threshold can be exactly
determined: (1) If two hyperedges both have pendant vertices at the
same boundary vertex, then consider each to have probability
$\sqrt{p}$ of being open, which is equivalent to one edge with
probability $p$. (2) If two hyperedges have two boundary vertices in
common and in both hyperedges these boundary vertices are adjacent,
then to get one edge with probability $p$, consider each edge to be
open with probability $s$ which satisfies $p = s^2 + 2 s (1-s) = 1 -
(1-s)^2$, which implies that $s = 1 - \sqrt{1-p}.$ Note that in both
cases these edge probability functions are increasing functions of
the parameter $p$, which is sufficient for the reasoning of this article to apply.

\section{Infinitely Many Structures}\label{equality-section}

In section 4, Ziff and Scullard \cite{Ziff+Scullard-JPhysA-2006}
illustrate a third self-dual hypergraph, in addition to the
triangular lattice structure and bow-tie lattice structure. We
provide another self-dual hypergraph, which has not appeared
previously, in Figure~\ref{Ziffnew}.

\begin{figure}
$$\includegraphics[scale=.50]{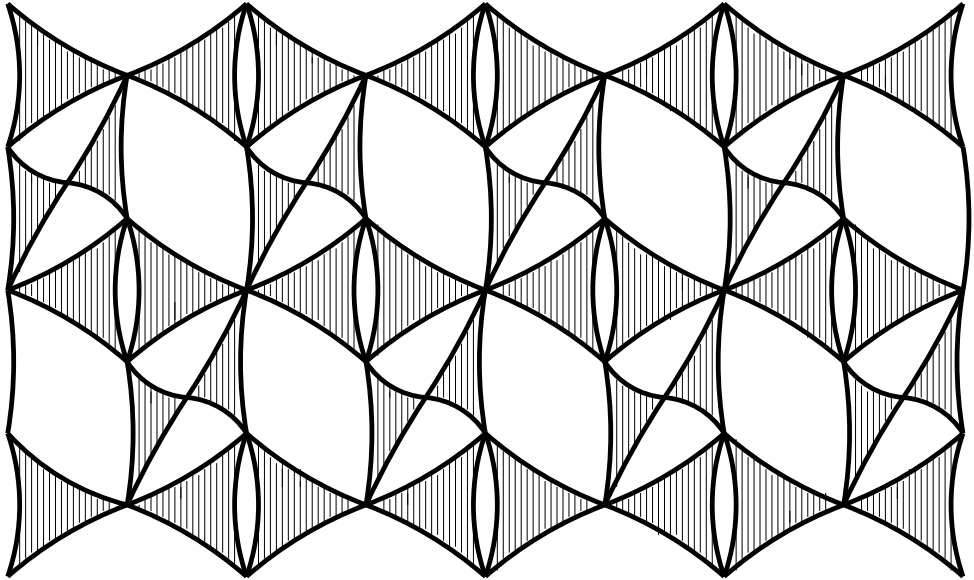}$$
\caption{A self-dual hypergraph.}
\end{figure}\label{Ziffnew}

Ziff and Scullard \cite{Ziff+Scullard-JPhysA-2006} mention that many other self-dual arrangements
can be constructed.  In this section, we make make this comment
more precise, by showing that there are in fact infinitely many self-dual hypergraphs.

Note that the bond percolation threshold of a graph constructed from
a generator and a self-dual hypergraph is uniquely
determined by the connectivity equation.  By filling all
self-dual 3-uniform hypergraphs with the same generator, we construct infinitely many
planar lattices which have the same value for the bond percolation
threshold.

\clearpage

\begin{figure}[pbht]
$$\includegraphics[scale=.50]{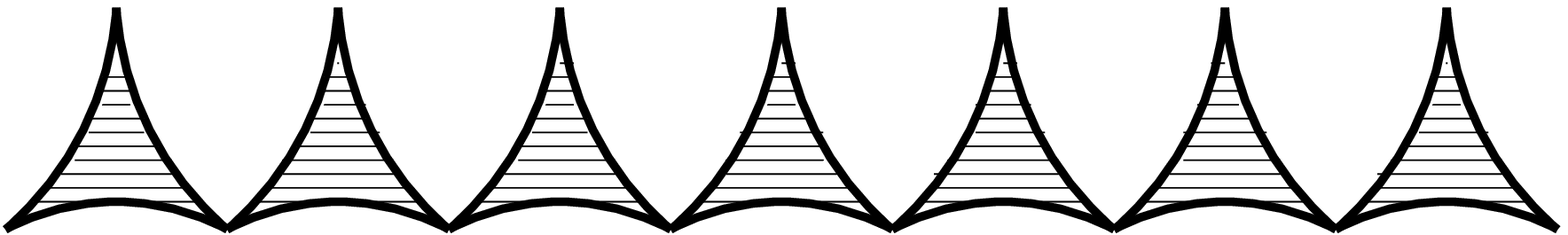}$$
\vspace{5mm}
$$\includegraphics[scale=.50]{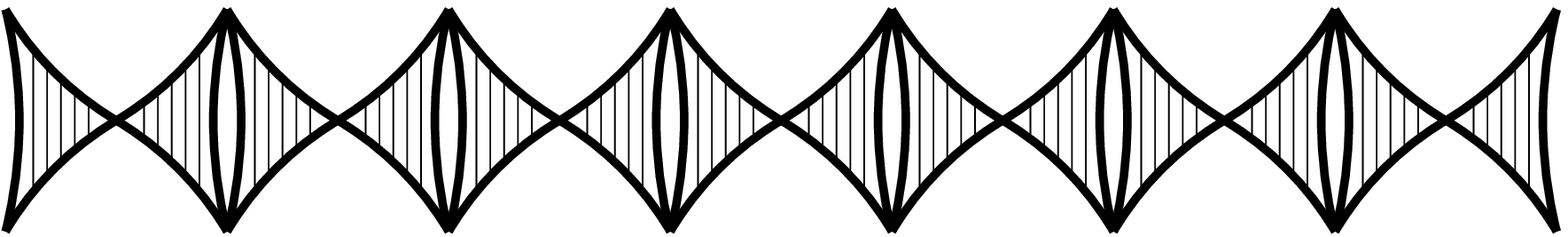}$$
\vspace{5mm}
$$\includegraphics[scale=.45]{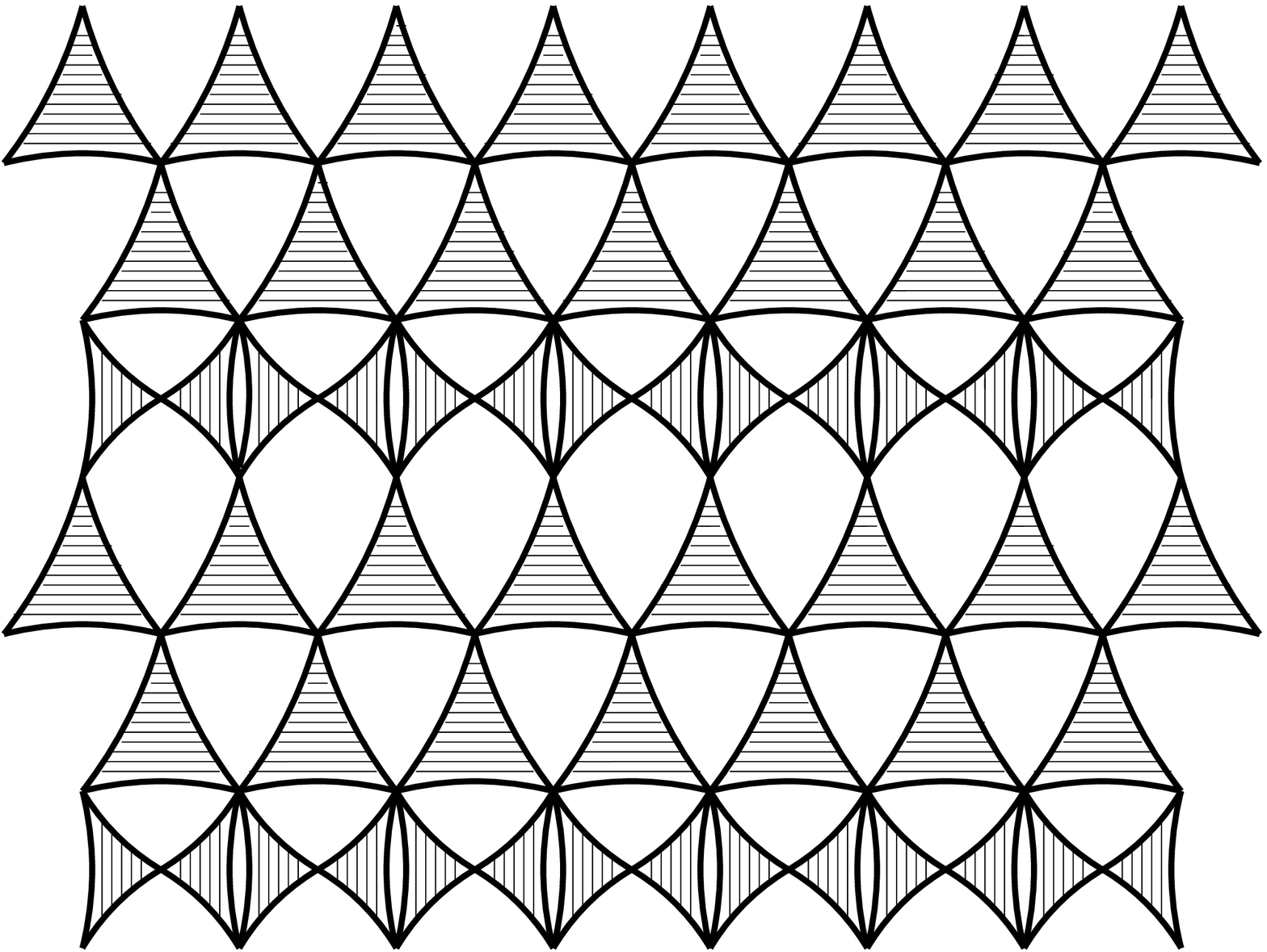}$$
 \caption{Top: A row of
upward-pointing triangles. Middle: A row of bow-ties. Bottom: A
periodic self-dual 3-regular hypergraph which alternates one row of
bow-ties with two rows of triangles.}
\end{figure}\label{strips}

We may construct self-dual hypergraphs using rows of
similarly-oriented triangles and rows of similarly-oriented
bow-ties, as shown in Figure 6.  To form an infinite
periodic self-dual hypergraph, we combine these into a periodic
pattern. Consider forming a sequence of rows which alternates between $m$
rows of triangles and $n$ rows of bow-ties, for any $m \geq 1$ and
$n \geq 1$.  When constructing the dual hypergraph,
each triangle is reflected 180 degrees, while the row of bow-ties is
translated so the boundary vertices match up with the triangle
vertices. As a result, the dual hypergraph is isomorphic to the
original hypergraph reflected and translated.  See
Figure 6 for an illustration.

We may construct another infinite collection of self-dual
hypergraphs similarly.  At the top of Figure~\ref{bent}, we show a row of
triangles with the top of every third triangle moved to the right.
Each time a top vertex is moved to the right, it produces two
hyperedges which share two boundary vertices, as occurs in the
bow-tie lattice structure. Consider forming a sequence of rows which
alternates between $m$ rows of triangles and $n$ rows of triangles
with tops moved, in a periodic manner.  Again, the dual hypergraph
is isomorphic to the original hypergraph  reflected and translated.
See Figure~\ref{bent}, which illustrates this construction. Note
that an arrangement in which all rows have every other top moved, the
hypergraph is isomorphic to the bow-tie hypergraph.

\begin{figure}[pbht]
$$\includegraphics[scale=.60]{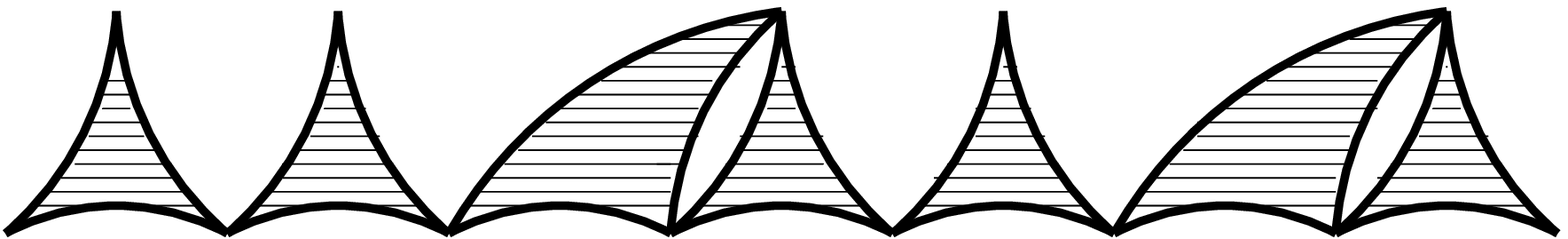}$$
\vspace{5mm}
$$\includegraphics[scale=.55]{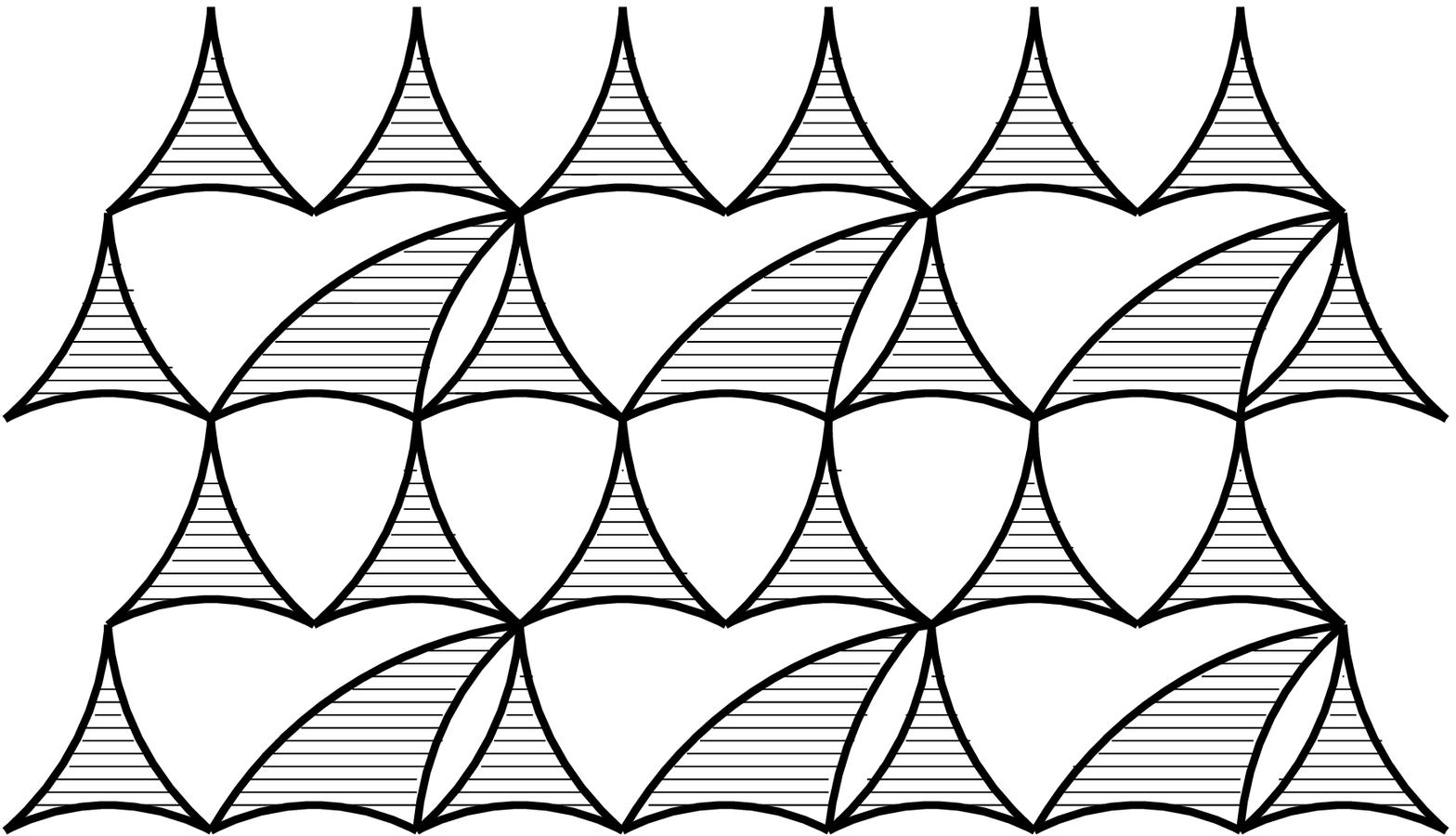}$$
\caption{Top: A row of triangles with every third top moved to the
right. Bottom: A self-dual hypergraph constructed by alternating
rows of triangles and rows of triangles with every other top moved
to the right.}
\end{figure}\label{bent}

\section{Infinitely-Many Threshold Values for Triangle-Dual
Structures}

\begin{figure}[pbht]
$$\includegraphics[scale=.28]{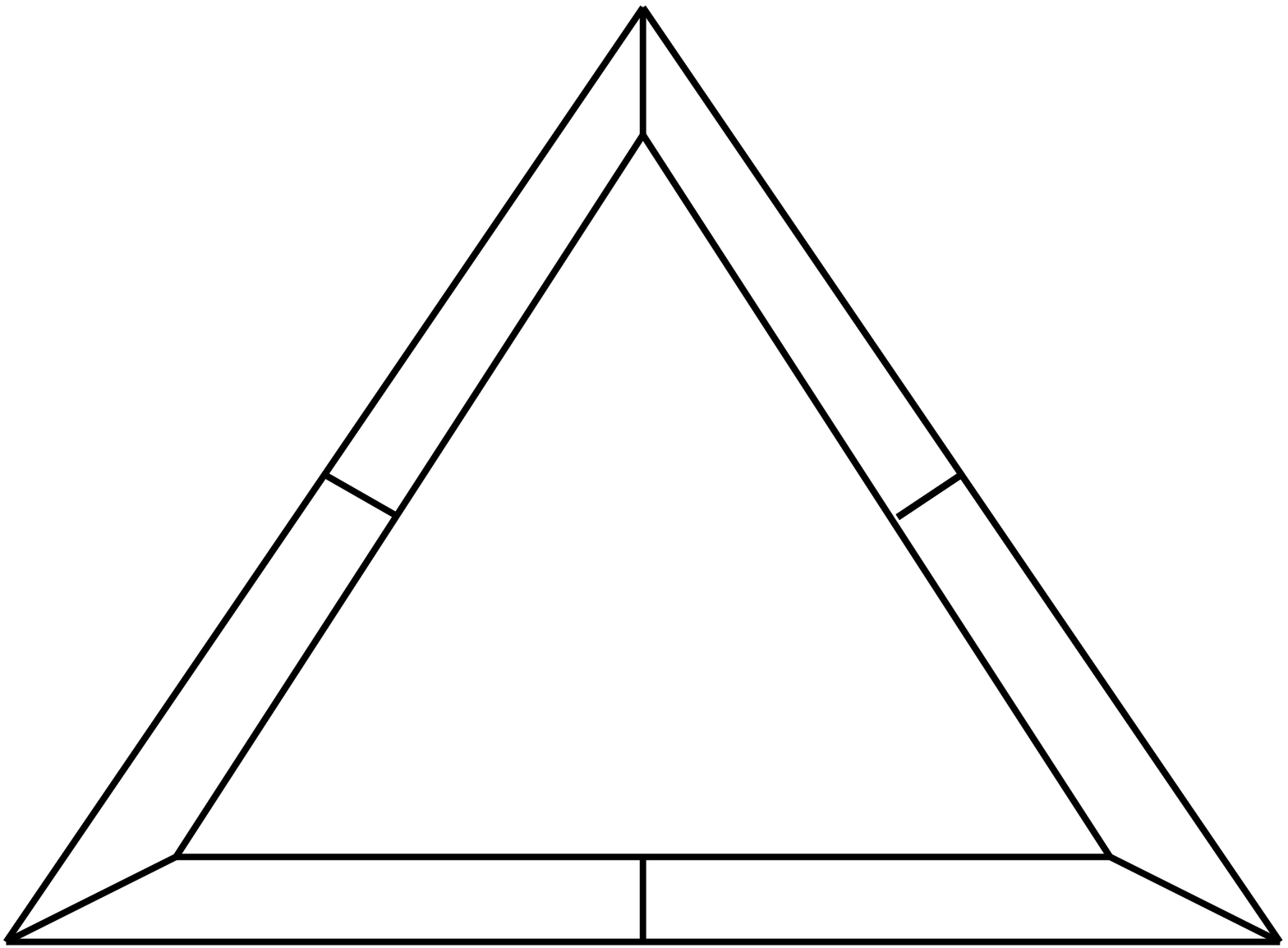} \hspace{10mm}
\includegraphics[scale=.28]{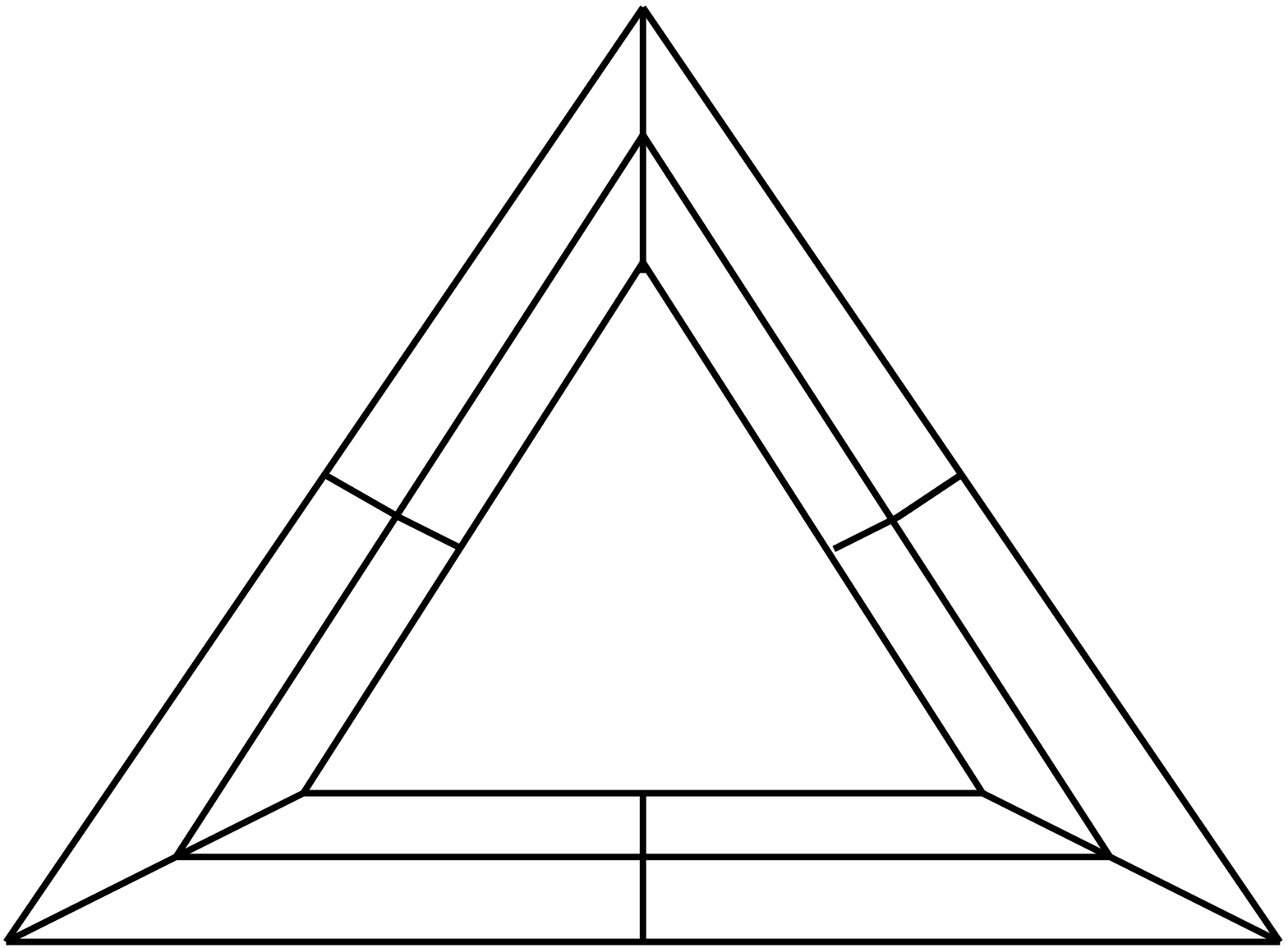}$$
$$\includegraphics[scale=.28]{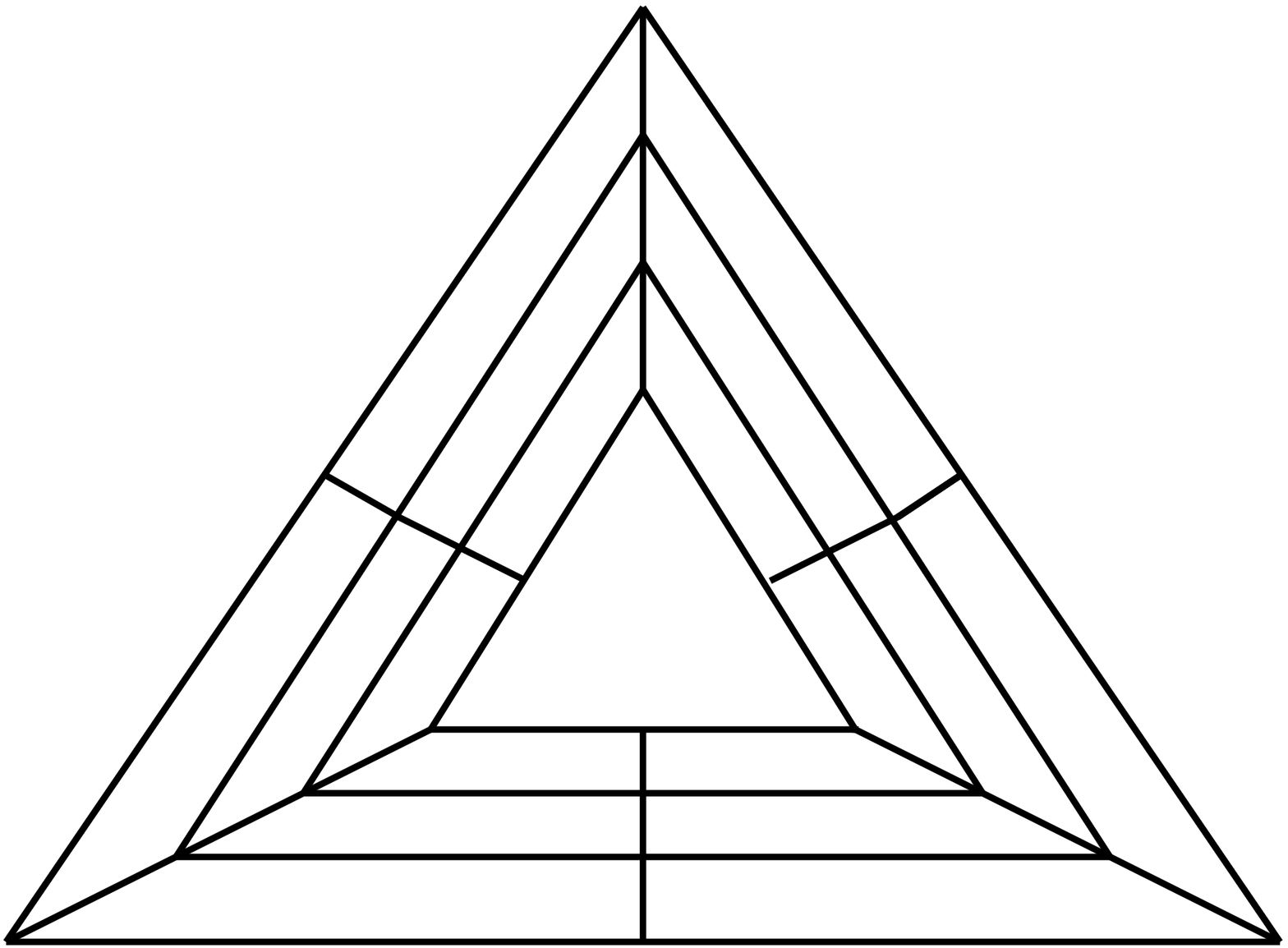} \hspace{10mm}
\includegraphics[scale=.28]{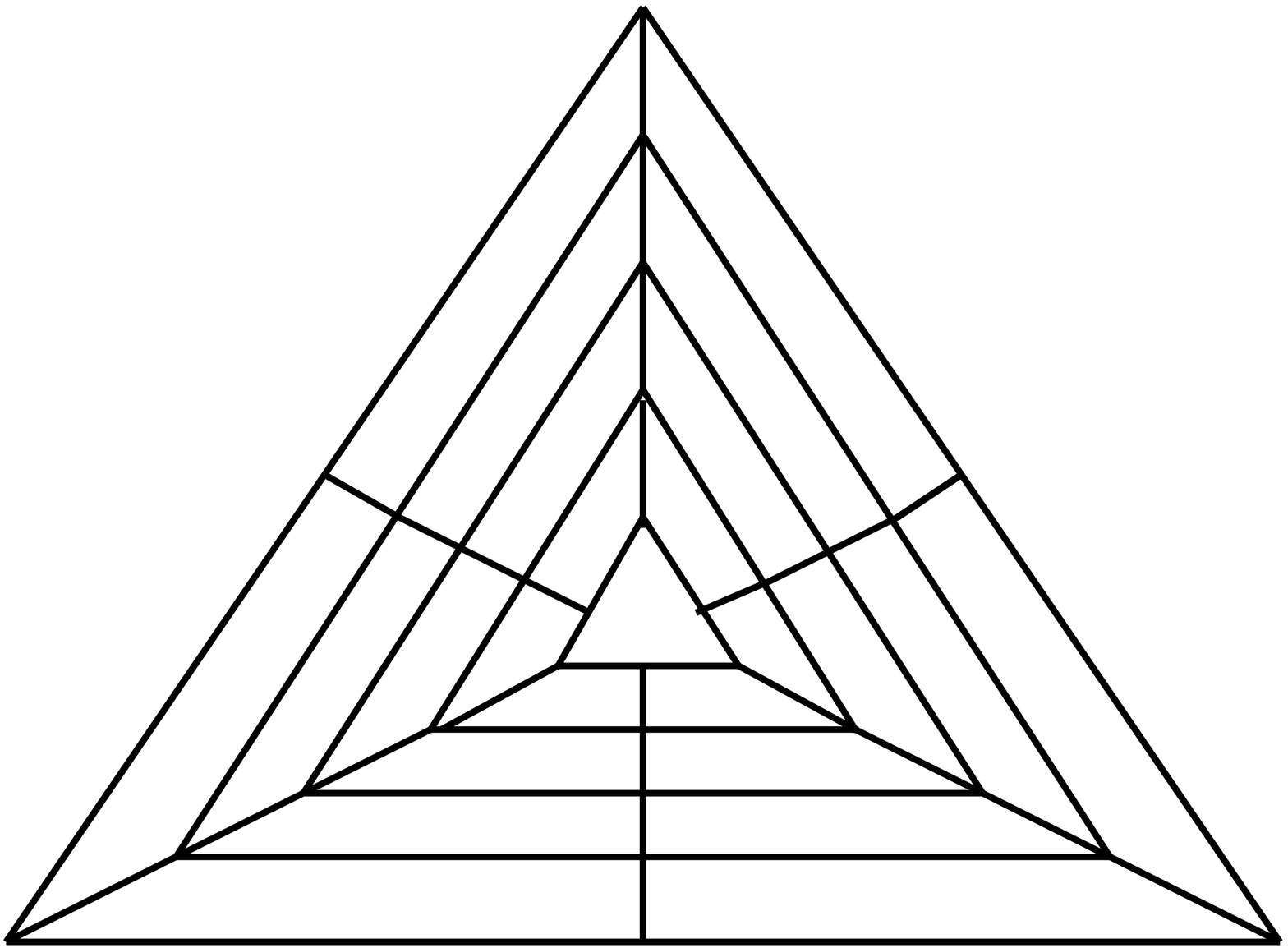}$$
\caption{The first four graphs in an infinitely sequence of nested
generators. Each contains all previous graphs in the sequence as
subgraphs.}
\end{figure}\label{nested-generators}

Figure~\ref{nested-generators} shows a sequence of generator graphs
that are nested so that each graph contains every previous graph as
a subgraph. Each generator can be put embedded in infinitely many
triangle-dual arrangements, producing infinitely many lattices which
have the same bond percolation threshold.  (Note that the
construction does not have single edges connecting the boundary
vertices, in order to avoid complications such as the lattice or its
dual lattice having split edges or subdivided edges.)  By the strict
subgraph relationships between the generators, a result of Aizenmann
and Grimmett \cite{Aizenman+Grimmett-1991} shows that the
percolation thresholds of the families of graphs for different
generators are not equal. Therefore, we have established that for
infinitely many real numbers $a$ there are infinitely many periodic
lattice graphs with bond percolation threshold equal to $a$.

Recall that by Fisher's \cite{Fisher-1961} bond-to-site
transformation, the bond percolation model on a lattice $L$ is
equivalent to the site percolation model on the covering lattice (in
the physics literature) or line graph (in the mathematical
literature), so that their percolation thresholds are equal.  
[Note that the result of the bond-to-site transformation is not always planar, so two-dimensional site models use the concept of ``matching lattice" rather than dual lattice.]
Thus,
we have also shown that for infinitely many values $a$ there are
infinitely many site percolation models with percolation threshold
equal to $a$.

\section{Concluding Remarks}

We have provided sufficient conditions on the underlying lattice for
correct application of the methods of Ziff and Scullard for
determining exact bond percolation thresholds.  The method applies
if the lattice is constructed from an infinite connected planar
periodic 3-regular hypergraph with one axis of symmetry, using a
generator which is a finite connected planar graph with three
boundary vertices. These conditions are naturally stated in terms of
planar hypergraphs, reflect the need for planarity of the lattice in
order to apply graph duality results, and reflect the need for
technical conditions for validity of the mathematical proofs.

In particular, note that sufficient conditions reflect the current
state of knowledge in mathematical percolation theory, requiring
periodicity and an axis of symmetry. It is plausible that the
results may be valid in a larger context, where these conditions are
relaxed. As evidence for this, in a recent breakthrough, Bollob\'as
and Riordan \cite{Bollobas+Riordan,Bollobas+Riordan-PTRF} exactly
determined the site percolation threshold in a certain continuum
percolation model which is not periodic or symmetric.

Ziff and Scullard also apply their approach to site percolation
models and correlated percolation models (either all open or all
closed). The formalism allows correlated bonds within each
generator. As in the case of bond percolation, care must be taken to
understand the conditions on the generator and the connections
between generators that produce a valid solution. Research on these
issues is in progress.

Wierman \cite{Wierman-1992} used the so-called ``substitution
method" with the star-triangle transformation to show that, if they
exist, the critical exponents are equal for the triangular and
hexagonal lattice bond percolation models, and that they are equal
for the bond percolation models on the bow-tie and its dual lattice.
Sedlock and Wierman \cite{Sedlock+Wierman} generalized this result to the class of lattices
identified in this article.


\vspace{3mm}
 \noindent
 {\bf Acknowledgements:}
 R M Ziff acknowledges
support from the National Science Foundation Grant No. DMS-0553487.
J C Wierman acknowledges financial support from Johns Hopkins
University's Acheson J. Duncan Fund for the Advancement of Research
in Statistics.  J. C. Wierman gratefully acknowledges the financial
support and stimulating environment of the Mittag-Leffler Institute
of the Swedish Royal Academy of Sciences during his sabbatical visit
in spring 2009.

%
%
%
%


\begin{thebibliography}{99}
\footnotesize

\bibitem{Aizenman+Grimmett-1991} Aizenman, M and Grimmett, G R
(1991)
Strict monotonicity for critical points in percolation and
ferromagnetic models.
{\em J. Stat. Phys.} {\bf 63,} 817--835.


\bibitem{Bollobas+Riordan} Bollob\'as, B and Riordan, O (2006).
{\em Percolation}. Cambridge University Press, Cambridge.

\bibitem{Bollobas+Riordan-PTRF} Bollob\'as, B and Riordan, O (2006)
The critical probability for random Voronoi percolation in the plane
is 1/2.
{\em Probabability Theory and Related Fields} {\bf 136,} 417--468.

\bibitem{BroadbentHammersley1957} Broadbent, S R and
Hammersley, J M (1957)
Percolation processes. I. Crystals and mazes.
{\em Proc. Camb. Phil. Soc.} {\bf 53,} 629--641.

\bibitem{Chayes+Lei-JStatPhys-2006} Chayes, L and Lei, H K (2006)
Random cluster models on the triangular lattice.
{\em J. Stat. Phys.} {\bf 122,} 647--670.

\bibitem{Fisher-1961} Fisher, M E (1961)
Critical probabilities for
cluster size and percolation problems.
{\em J. Math. Phys.} {\bf 2,} 620--627.

\bibitem{Grimmett1999} Grimmett, G R (1999). {\em Percolation}. Springer.

\bibitem{Hughes-volume2} Hughes, B (1996). {\em Random Walks and Random
Environments. Volume 2: Random Environments}. Clarendon
Press, Oxford.

\bibitem{Kesten1980} Kesten, H (1980)
The critical probability of bond percolation
on the square lattice equals 1/2.
{\em Comm. Math. Phys.} {\bf 74,} 41--59.

\bibitem{Kesten1982} Kesten, H (1982) {\em Percolation Theory for Mathematicians}. Birkh\"auser, Boston.

\bibitem{Lyons1990} Lyons, R (1990)
Random walks and percolation on tree.
{\em Ann. Probab.} {\bf 18,} 931--958.

\bibitem{May+Wierman-2003} May, W D and Wierman, J C (2003)
Recent improvements to
the substitution method for bounding percolation thresholds.
{\em Congressus Numerantium} {\bf 162,} 5--25.

\bibitem{May+Wierman-2005} May, W D and Wierman, J C (2005)
Using symmetry to
improve percolation threshold bounds.
{\em Combinatorics, Probabability and Computing} {\bf 14,} 549--566.

\bibitem{Sahimi1994}  Sahimi, M (1994)  {\em Applications of
Percolation Theory}. Taylor \& Francis.

\bibitem{Scullard-PRE-2006} Scullard, C R (2006)
Exact site percolation thresholds using a site-to-bond
transformation and the star-triangle transformation.
{\em Phys. Rev. E} {\bf 73,} 016107.

\bibitem{Scullard+Ziff-PRE-2006} Scullard, C R and Ziff, R M (2006)
Predictions of bond percolation thresholds for the
kagom\'e and Archimedean (3,$12^2$) lattices.
{\em Phys. Rev. E} {\bf 73,} 045102(R)

\bibitem{Sedlock+Wierman} Sedlock, M R A and Wierman, J C (2009) Equality of bond-percolation critical exponents for pairs of dual lattices. {\em Physical Review E} {\bf 79,} 05119.

\bibitem{Smythe+Wierman-1978} Smythe, R T and Wierman, J C (1978)
{\em First-Passage Percolation on the Square Lattice.} (Lecture Notes
in Mathematics, {\bf 671.}) Springer, Berlin.

\bibitem{StaufferAharony1991} Stauffer, D and Aharony, A
(1991) {\em Introduction to Percolation Theory}. Taylor \& Francis.

\bibitem{Wierman1981}  Wierman, J C (1981)
Bond percolation on the honeycomb
and triangular lattices.
{\em Adv. Appl. Probab.} {\bf 13,} 298--313.

\bibitem{Wierman1984-1525} Wierman, J C (1984)
A bond percolation critical
probability determination based on the star-triangle transformation.
{\em J. Phys. A: Math. Gen.} {\bf 17,} 1525--1530.

\bibitem{Wierman-1990} Wierman, J C (1990)
Bond percolation critical probability
bounds for the Kagome lattice by a substitution method.
{\em Disorder in Physical Systems,} Oxford University Press, 349--360.

\bibitem{Wierman-1992} Wierman, J C (1992)
Equality of the bond percolation critical
exponents for two pairs of dual lattices,
{\em Combinatorics, Probab. Comput.} {\bf 1,} 95--105.

\bibitem{Wierman-1995} Wierman, J C (1995)
Substitution method critical probability
bounds for the square lattice site percolation model.
{\em Combinatorics, Probability and Computing,} {\bf 4,} 181--188.

\bibitem{Wierman-2001} Wierman, J C (2001)
Site percolation critical probability
bounds for the $(4,8^2)$ and $(4,6,12)$ lattices.
{\em Congressus Numerantium} {\bf 150,} 117--128.

\bibitem{Wierman-2002-RSA} Wierman, J C (2002)
Bond percolation critical probability
bounds for three Archimedean lattices.
{\em Rand. Struct. \& Alg.} {\bf 20,} 507--518.

\bibitem{Wierman-2002-CPC} Wierman, J C (2002)
An improved upper bound for the hexagonal
lattice site percolation critical probability.
{\em Combinatorics, Probab. Comput.} {\bf 11,} 629--643.

\bibitem{Wierman-HICStat} Wierman, J C (2006)
Construction of infinite
self-dual graphs.
{\em Proc. 5th Hawaii International Conference on Statistics,
Mathematics and Related Fields,} [CD-ROM]

\bibitem{Wu-PRL-2006} Wu, F Y (2006) New critical frontiers for the Potts and percolation models. {\em Physical Review Letters} {\bf 96}, 090602.

\bibitem{Ziff-PRE-2006} Ziff, R M (2006)
Generalized
cell--dual-cell transformation and exact thresholds for percolation.
{\em Phys. Rev. E} {\bf 73,} 016134.

\bibitem{Ziff+Scullard-JPhysA-2006} Ziff, R M and Scullard, C R (2006)
Exact bond percolation thresholds in two dimensions.
{\em J. Phys. A: Math. Gen.} {\bf 39,} 15083--15090.


\end{thebibliography}
\end{document}